\documentclass[a4paper,twocolumn,11pt,accepted=2026-07-20]{quantumarticle}
\pdfoutput=1
\usepackage[utf8]{inputenc}
\usepackage[english]{babel}
\usepackage[T1]{fontenc}
\usepackage{amsmath}
\usepackage{amssymb}
\usepackage{hyperref}
\usepackage{tikz}
\usepackage{enumitem}
\usepackage[numbers,sort&compress]{natbib}

\newcommand{\Tr}[0]{\text{Tr}}

\newcommand{\R}[1]{\textrm{#1}}
\newcommand{\Cs}{{}^{13}\R{C}}
\newcommand{\mH}[0]{\mathcal{H}}

\newcommand{\ket}[1]{\left\vert{#1}\right\rangle}

\begin{document}
\title{Modulator-Assisted Zeno Control of Energy Transfer in Quantum Batteries}
\author{Songbo Xie}
\affiliation{Department of Electrical and Computer Engineering, North Carolina State University, Raleigh, North Carolina 27606, USA}
\orcid{0000-0002-9136-9481}
\author{Manas Sajjan}
\affiliation{Department of Electrical and Computer Engineering, North Carolina State University, Raleigh, North Carolina 27606, USA}
\orcid{0000-0001-7436-5422}
\author{Ashok Ajoy}
\affiliation{Department of Chemistry, University of California, Berkeley, Berkeley, California 94720, USA}
\orcid{0000-0003-3242-2913}
\author{Sabre Kais}
\email{skais@ncsu.edu}
\affiliation{Department of Electrical and Computer Engineering, North Carolina State University, Raleigh, North Carolina 27606, USA}
\affiliation {Miller Institute for Basic Research in Science
468 Donner Lab, Berkeley, California 94720, USA}
\orcid{0000-0003-0574-5346}
\maketitle

\begin{abstract}
  Efficient operation of quantum batteries requires not only fast energy transfer but also the ability to halt the charging process to prevent reverse flow. Existing approaches typically rely on direct control of the charger-battery interaction, which can be experimentally demanding. Here we propose a modulator-assisted quantum battery protocol that enables indirect control of energy transfer while keeping the interaction always on. By applying repeated local unitary operations to an auxiliary modulator qubit, we exploit a Zeno-like mechanism to dynamically reshape the effective Hamiltonian and switch the charger-battery coupling on and off. We demonstrate this mechanism in a minimal three-body model and show that it remains effective beyond the ideal fast-control limit. We further extend the protocol to a collective many-body architecture, where it preserves the characteristic enhancement of charging power, scaling as $N^{3/2}$ with the number of battery units. We also discuss a possible implementation in an NV-$\Cs$ spin platform. Our results establish modulator-assisted Zeno control as a scalable route to regulating energy transfer in quantum batteries. 
\end{abstract}

\section{Introduction}

\begin{figure}[!b]
    \centering
    \includegraphics[width=0.48\textwidth]{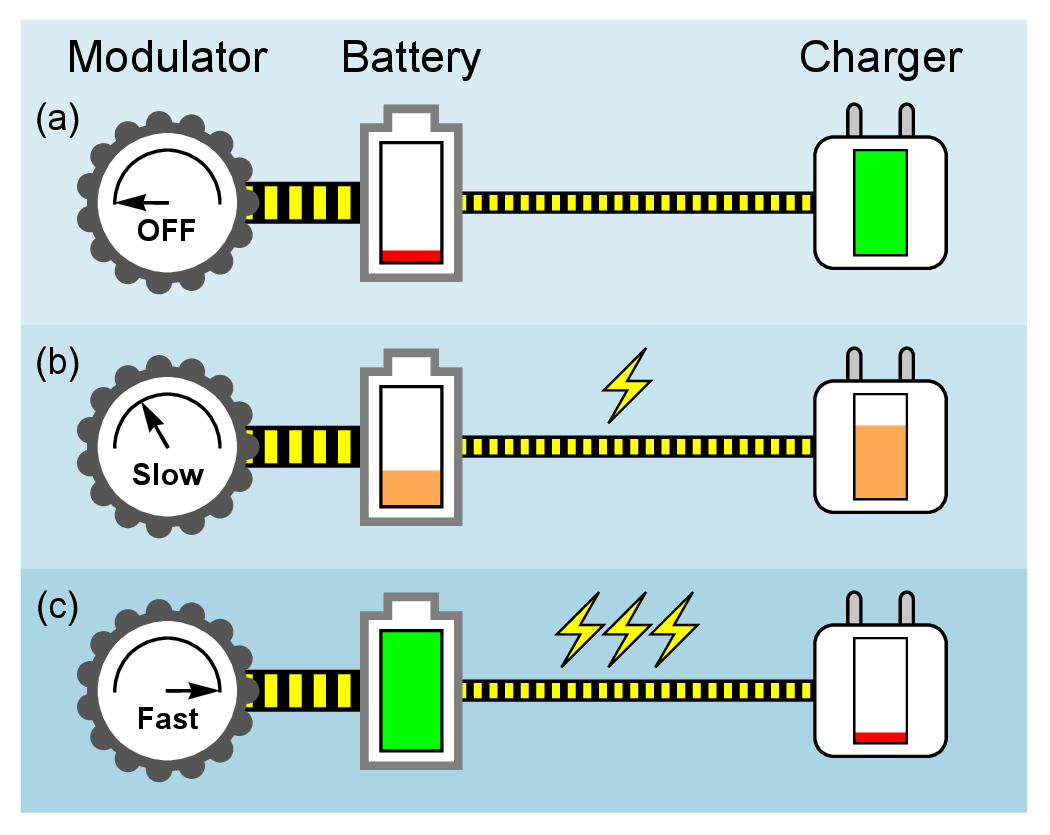}
    \caption{A schematic representation of the modulator-assisted quantum battery, where the battery-charger coupling is always on. The effective coupling strength is modified through adding kicks to an external modulator, which only interacts locally with the battery.}
    \label{fig:kcb}
\end{figure}

Quantum batteries have attracted growing attention as a platform for exploring how quantum effects can enhance the storage and transfer of energy at small scales \cite{campaioli2024colloquium,alicki2013entanglement,friis2018precision,santos2019stable,dou2022highly,quach2022superabsorption,hu2022optimal,maillette2023experimental}. A central focus in this field has been the identification of charging protocols that outperform their classical counterparts, particularly through collective many-body effects that can enhance charging power~\cite{hovhannisyan2013entanglement,binder2015quantacell,campaioli2017enhancing,rossini2020quantum,andolina2019quantum,julia2020bounds,gyhm2022quantum,gyhm2024beneficial,andolina2025genuine}. In practice, however, a useful charging protocol must accomplish more than rapid energy transfer alone: it must also provide a mechanism to halt the transfer once the battery is charged, in order to prevent reverse energy flow and preserve the stored energy. The ability to switch between charging and storing phases is therefore a fundamental operational requirement for quantum batteries.

In most existing protocols, this control is achieved by directly tuning the charger-battery interaction or by detuning the relevant subsystems in time \cite{niskanen2006tunable,niskanen2007quantum,brennen1999quantum,zeiher2016many,bluvstein2021controlling,sorensen1999quantum,leibfried2003experimental}. Such approaches are effective in idealized settings and have motivated a broad class of tunable-coupling and quench-based charging schemes \cite{andolina2018charger,ferraro2018high,farina2019charger,crescente2020ultrafast,delmonte2021characterization}. However, they generally rely on direct access to the primary energy-transfer subsystem, requiring local control over the charger, the battery, or the interaction link itself. These requirements become increasingly restrictive in architectures where the energy-storing subsystem is spatially separated, where direct intervention is experimentally costly, or where repeated control on the battery may introduce unwanted back-action, additional hardware overhead, and noise.

This challenge becomes even more pronounced in many-body quantum batteries. While collective charging architectures can provide enhanced charging power, they also increase the complexity of control, since one must regulate energy flow across an extended interacting system without destroying the collective advantage. This raises a natural question: \textit{Can one control the charging and storing process of quantum batteries indirectly, through operations on a small auxiliary subsystem only, while leaving the charger-battery interaction always on and preserving scalability to collective architectures?}

In this work, we answer this question affirmatively by introducing a modulator-assisted control paradigm for charger-mediated quantum batteries. Our scheme augments the battery system with an auxiliary two-level system---the modulator---which is coupled locally to the battery subsystem but is itself the only degree of freedom directly controlled. By applying repeated local unitary kicks to the modulator, we exploit a Zeno-like mechanism to dynamically reshape the effective Hamiltonian of the coupled system \cite{facchi2008quantum}. As a result, the effective charger-battery coupling can be switched on and off without ever directly tuning the charger-battery interaction itself. See Figure~\ref{fig:kcb} for a schematic representation.

This use of the quantum Zeno effect goes beyond its conventional role as a tool for freezing the state of a directly controlled subsystem. Here, local control of the modulator is converted into indirect control of a remote energy-transfer channel, enabling nonlocal regulation of charging and storage while the physical interaction remains constantly present \cite{xie2025strong}. In this way, the protocol establishes a distinct control architecture for quantum batteries based on indirect, local, and always-on interaction control.

We develop this framework at two levels. First, we analyze a minimal three-body setting consisting of a charger qubit, a battery qubit, and a modulator qubit, and show that the protocol enables a complete charging-storage cycle without direct intervention on the charger-battery pair. We further examine the finite-kicking regime and show that the protocol remains effective beyond the ideal dense-kicking limit. Second, we extend the protocol to a collective many-body architecture in which a single cavity mode charges an array of $N$ battery qubits under the control of a single modulator. In this setting, we show that the charging process remains suppressed in the absence of control, is restored by local kicks on the modulator, and retains the characteristic collective charging advantage, with maximal charging power scaling as \(N^{3/2}\).

These results establish modulator-assisted Zeno control as a scalable route to indirect charging regulation in quantum batteries, and suggest a general strategy for controlling energy flow in composite quantum systems without direct access to the primary energy-storage degrees of freedom. We further discuss how the current indirect-control paradigm may be implemented in an experimentally motivated NV-$\Cs$ spin platform.

\section{Minimal Modulator-Assisted Quantum Battery}

Before extending to the many-body setting, we begin with a minimal charger-mediated quantum-battery model consisting of a single battery qubit \(b\) coupled to a charger qubit \(c\). To implement the indirect-control paradigm, we introduce an additional modulator qubit \(m\), which interacts only with the battery \(b\). The modulator could be coupled to the charger equivalently, but here we choose the battery. The modulator and battery together form an ``expanded battery'' system, described by the Hamiltonian (\(\hbar=1\))
\begin{equation}\label{hmc}
    \mathcal{H}_{mb} = -\dfrac{\omega_0}{2} \sigma_m^z + \dfrac{J}{2} \left(\sigma_m^x \sigma_b^x + \sigma_m^y \sigma_b^y\right) - \dfrac{\omega_0}{2} \sigma_b^z.
\end{equation}
For maximal controllability, we take the modulator and battery to be resonant, both with frequency \(\omega_0\), and \(J\) is their coupling strength, assumed to satisfy \(J<\omega_0\). The subscripts \(\{m,b,c\}\) label the qubits, and the superscripts \(\{x,y,z\}\) label the Pauli operators:
\(\sigma^x = |1\rangle\langle0| + |0\rangle\langle1|\),
\(\sigma^y = i|1\rangle\langle0| - i|0\rangle\langle1|\), and
\(\sigma^z = |0\rangle\langle0| - |1\rangle\langle1|\).

The eigenvectors and eigenvalues of \(\mathcal{H}_{mb}\) are
\begin{equation}
\begin{aligned}
|v_0\rangle &= |0\rangle_m|0\rangle_b, 
&\ \text{with }\lambda_0 &= -\omega_0,\\
|v_1\rangle &= \tfrac{1}{\sqrt{2}}\bigl(|0\rangle_m|1\rangle_b-|1\rangle_m|0\rangle_b\bigr), 
&\ \text{with }\lambda_1 &= -J,\\
|v_2\rangle &= \tfrac{1}{\sqrt{2}}\bigl(|0\rangle_m|1\rangle_b+|1\rangle_m|0\rangle_b\bigr), 
&\ \text{with }\lambda_2 &= J,\\
|v_3\rangle &= |1\rangle_m|1\rangle_b, 
&\ \text{with }\lambda_3 &= \omega_0.
\end{aligned}
\end{equation}

The expanded battery \(mb\) is coupled to the charger \(c\) through the Hamiltonian
\begin{equation}\label{hmbc}
    \begin{split}
        &\mathcal{H}_{mbc}=\mathcal{H}_{mb}+\mathcal{V}_{bc}+\mathcal{H}_c,\\
        \text{with}\quad &\mathcal{V}_{bc}=\dfrac{g}{2}(\sigma^{x}_b\sigma^{x}_c+\sigma^{y}_b\sigma^{y}_c),\\
        \text{and}\quad &\mathcal{H}_c=-\dfrac{\omega_1}{2}\sigma_c^z.
    \end{split}
\end{equation}
Here, \(g\) is the coupling strength between \(b\) and \(c\), which we assume to be weak, \(g\ll J\). The charger frequency \(\omega_1\) will be specified below. We define the battery energy $E_b$ and the charger energy $E_c$ as
\begin{equation}
    \begin{split}
        E_b=&\Tr[\rho(\mathcal{H}_{mb}+\omega_0\mathbb{I}_{mb})],\\
        E_c=&\Tr[\rho(\mathcal{H}_c+\tfrac{\omega_1}{2}\mathbb{I}_c)],
    \end{split}
\end{equation}
Here, \(\rho\) is the density operator of the three-qubit system, and \(\mathbb{I}\) is the identity operator. These constant energy shifts ensure that both the battery and the charger have zero ground-state energy, without affecting the dynamics.

We initialize the system in the product state \( |v_0\rangle_{mb} |1\rangle_c\), where the expanded battery is in its ground state and the charger is in its excited state, carrying the energy to be transferred. In the eigenbasis of the expanded battery, as shown in Appendix A, the coupling \(\mathcal{V}_{bc}\) in Eq.~\eqref{hmbc} can be rewritten as
\begin{equation}\label{vbc}
    \begin{aligned}
        \mathcal{V}_{bc}
        = \dfrac{g}{\sqrt{2}} \Big[ \,
        &\big(
        |v_1\rangle\langle v_0|_{mb}
        + |v_2\rangle\langle v_0|_{mb}
        + |v_3\rangle\langle v_2|_{mb} \\
        &\quad - |v_3\rangle\langle v_1|_{mb}
        \big)\otimes |0\rangle\langle 1|_c
        + \mathrm{H.c.}
        \Big].
    \end{aligned}
\end{equation}
Here, H.c.~denotes the Hermitian conjugate. Starting from the initial state, only two transitions are allowed: \( |v_0\rangle \leftrightarrow |v_1\rangle \), with energy difference \( \omega_0-J \), and \( |v_0\rangle \leftrightarrow |v_2\rangle \), with energy difference \( \omega_0+J \). For the charger, if we choose \(\omega_1=\omega_0\), the transition \( |0\rangle \leftrightarrow |1\rangle \) has energy difference \( \omega_0 \). The charger is therefore detuned from both allowed expanded-battery transitions by \(|\Delta_{bc}|=J\). Since \(J\gg g\), this detuning strongly suppresses the interaction Eq.~\eqref{vbc}, and energy exchange is effectively blocked. Moreover, because all three qubits are initially prepared in eigenstates, they remain unchanged in these eigenstates in the absence of resonant exchange.

To activate energy transfer, we freeze the modulator in its initial state \( |0\rangle \) so as to induce a Zeno-like control. In principle, this could be achieved by repeated projective measurements. However, projective measurements produce stochastic outcomes, and the probability of maintaining the desired state decreases exponentially with the number of repetitions. Instead, we adopt a bang-bang control protocol \cite{dhar2006preserving,singh2014experimental}, implemented through repeated local unitary kicks \(\sigma_m^z\) applied to the modulator qubit \(m\).

This control is not introduced as a time-dependent term in the Hamiltonian. Rather, it is realized as a stroboscopic sequence of free evolutions under \(\mathcal{H}_{mbc}\), interleaved with unitary kicks. The corresponding evolution operator is
\begin{equation}
    U(\tau)=\sigma_m^z e^{-i\mathcal{H}_{mbc}\tau} \sigma_m^z e^{-i\mathcal{H}_{mbc}\tau}\cdots \equiv e^{-i\mathcal{H}_{mbc}^\text{eff}\tau}.
\end{equation}
As shown in Appendix B, the leading order in the Baker-Campbell-Hausdorff (BCH) expansion gives
\begin{equation}
    \mathcal{H}_{mbc}^\text{eff}=-\dfrac{\omega_0}{2}\sigma_m^z-\dfrac{\omega_0}{2}\sigma_b^z+\dfrac{g}{2}(\sigma^x_b\sigma^x_c+\sigma^y_b\sigma^y_c)-\dfrac{\omega_1}{2}\sigma_c^z,
\end{equation}
in which the \(mb\) coupling is completely cancelled. As a result, the modulator remains locked in its initial ground state, achieving the desired freezing effect.

While the Zeno effect is usually discussed as a mechanism for freezing the state of the directly controlled qubit, a less emphasized consequence is that it can also reshape the effective Hamiltonian of other qubits not directly acted upon by the unitary kicks \cite{xie2025strong}. Once the modulator is frozen in \(|0\rangle\), the remaining battery-charger subsystem is governed by
\begin{equation}\label{hbc}
    \begin{split}
        \mathcal{H}_{bc}=- \dfrac{\omega_0}{2} \sigma^z_b+g(\sigma^{+}_b\sigma^{-}_c+\sigma^{-}_b\sigma^{+}_c) - \dfrac{\omega_1}{2} \sigma^z_c,
    \end{split}
\end{equation}
where we use the standard notation \(\sigma^+=|1\rangle\langle0|\) and \(\sigma^-=|0\rangle\langle1|\).

With the above choice \(\omega_1=\omega_0\), the battery becomes resonant with the charger, and the interaction is restored. Since the charger is initialized in the excited state \(|1\rangle\) and the battery in the ground state \(|0\rangle\), energy is transferred from the charger to the battery under \(\mathcal{H}_{bc}\). The battery reaches its maximal excitation at time \(\pi/(2g)\).

\begin{figure}[!t]
    \centering
    \includegraphics[width=0.48\textwidth]{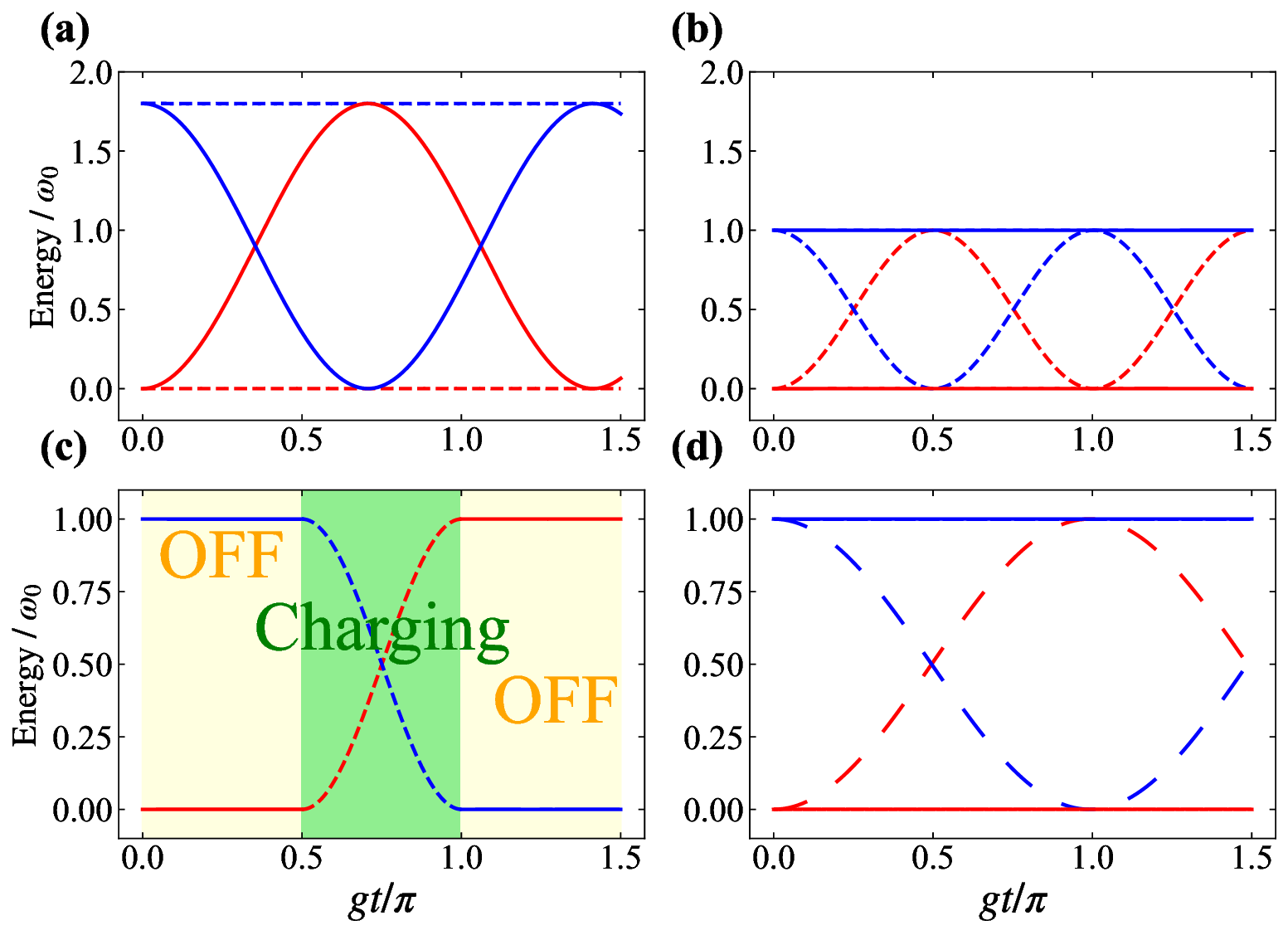}
    \caption{Plot legends: blue curves represent the charger energy \(E_c\), and red curves represent the battery energy \(E_b\). Solid lines indicate that no unitary kicks are applied, while dashed lines indicate that unitary kicks are applied to the modulator. We choose \(g=0.01\omega_0\) and \(J=0.8\omega_0\). {\bf Panel~(a)}: \(\omega_1 = \omega_0+J\) with kick interval \(\tau = \pi/(1000g)\); energy exchange is enabled when the kicks are off and disabled when the kicks are on. {\bf Panel~(b)}: \(\omega_1 = \omega_0\) with the same \(\tau\) as in (a); energy exchange is enabled when the kicks are on and disabled when they are off. {\bf Panel~(c)}: the same \(\omega_1\) and \(\tau\) as in (b); a charging cycle---consisting of an idle phase, followed by a charging phase, and then a storing phase---is achieved by selectively applying or not applying the kicks to the modulator. {\bf Panel~(d)}: the same \(\omega_1\) as in (b), but with a larger kick interval \(\tau = \pi/(100g)\). This leads to a slower charging speed while maintaining the same charging capacity.}
    \label{fig:dynamics}
\end{figure}

\section{Charging and Storage Control}

The main results are presented in Figure~\ref{fig:dynamics}. The blue curves show the charger energy \(E_c\), while the red curves show the battery energy \(E_b\). Solid lines correspond to the evolution without repeated unitary kicks \(\sigma_m^z\), whereas dashed lines correspond to the evolution with repeated unitary kicks applied to the modulator.

To verify Eq.~\eqref{vbc}, we first set \(\omega_1 = \omega_0+J\), so that the charger is resonant with the bare transition \( |v_0\rangle \leftrightarrow |v_2\rangle \) of the expanded battery when no unitary kicks are applied. We choose \(J=0.8\omega_0\). Under this resonance condition, energy exchange is shown by the solid curves in Figure~\ref{fig:dynamics}(a). The maximal energy transfer occurs at \(t = \pi/(\sqrt{2}g)\), consistent with the effective coupling strength \(g/\sqrt{2}\) in Eq.~\eqref{vbc}. The maximal transferred energy is \(\omega_0+J\).

By contrast, when repeated unitary kicks \(\sigma_m^z\) are applied to the modulator, energy exchange is suppressed. Let \(\tau\) denote the time interval between consecutive kicks, which we set to \(\tau = \pi/(1000g)\). These kicks remove the resonance condition and suppress energy transfer, as shown by the dashed curves in Figure~\ref{fig:dynamics}(a).

In Figure~\ref{fig:dynamics}(b), we instead set \(\omega_1 = \omega_0\), reversing the resonance condition. In this case, the charger and battery are off resonant when no unitary kicks are applied, and energy transfer is suppressed, as indicated by the flat solid curves. In contrast, when repeated unitary kicks are applied to the modulator, with the interval still chosen as \(\tau = \pi/(1000g)\), the system becomes resonant and energy transfer is restored, as shown by the dashed curves. The maximal transferred energy is \(\omega_0\), reached at \(t = \pi/(2g)\), consistent with the coupling strength \(g\) in Eq.~\eqref{hbc}.

From now on, we fix \(\omega_1 = \omega_0\). The above results show that the interaction between the charger and the battery can be switched on and off dynamically by controlling whether repeated unitary kicks are applied to the modulator. In Figure~\ref{fig:dynamics}(c), the protocol begins with an idle phase, indicated by the first yellow region, during which the energy remains stored in the charger. Next, repeated unitary kicks are applied to the modulator, activating the charger-battery interaction. This charging phase is marked by the green region and lasts for a duration of \(\pi/(2g)\), during which energy is maximally transferred from the charger to the battery. The kicks are applied at intervals of \(\tau = \pi/(1000g)\). Finally, to store the energy in the battery, the unitary kicks are turned off, thereby deactivating the interaction, as shown in the second yellow region. This sequence completes one full cycle of the charging protocol.

\textit{Tuning the kicking interval.---}In practice, achieving perfectly dense kicking is challenging. It is therefore important to examine how the system behaves as the kicking interval \(\tau\) increases. In the limit \(\tau \rightarrow 0\) (i.e., perfectly dense kicking), the effective detuning between the charger and battery vanishes. Conversely, when \(\tau \rightarrow \infty\) (i.e., no kicking), the detuning reaches \(|\Delta_{bc}|=J\). For finite values of \(\tau\) satisfying \(0 < \tau < \infty\), one might na\"ively expect the detuning to be an increasing function of \(\tau\) between these two extremes. According to the Jaynes-Cummings model, an increase in detuning leads to an increase in the Rabi frequency, suggesting a higher charging speed \cite{gerry2023introductory}. If the detuning indeed increases with \(\tau\), then the charging speed, which we define as \(1/T\)---where \(T\) is the time at which the battery first becomes maximally charged---would also be expected to increase with \(\tau\).

However, this expectation does not hold. In fact, for general finite values of \(\tau\) with \(0 < \tau < \infty\), a local effective Hamiltonian for the charger cannot be well defined, nor can the notion of detuning. In Figure~\ref{fig:dynamics}(d), we consider \(\tau = \pi/(100g)\), which is 10 times larger than that in Figure~\ref{fig:dynamics}(b). Contrary to the na\"ive expectation, the charging speed decreases with \(\tau\) rather than increases, while the amount of transferred energy remains unchanged at \(\omega_0\). Specifically, complete energy transfer now occurs at \(gt/\pi \approx 1\), in contrast to \(gt/\pi = 0.5\) in the limit \(\tau \rightarrow 0\).

\begin{figure}[!t]
    \centering
    \includegraphics[width=0.45\textwidth]{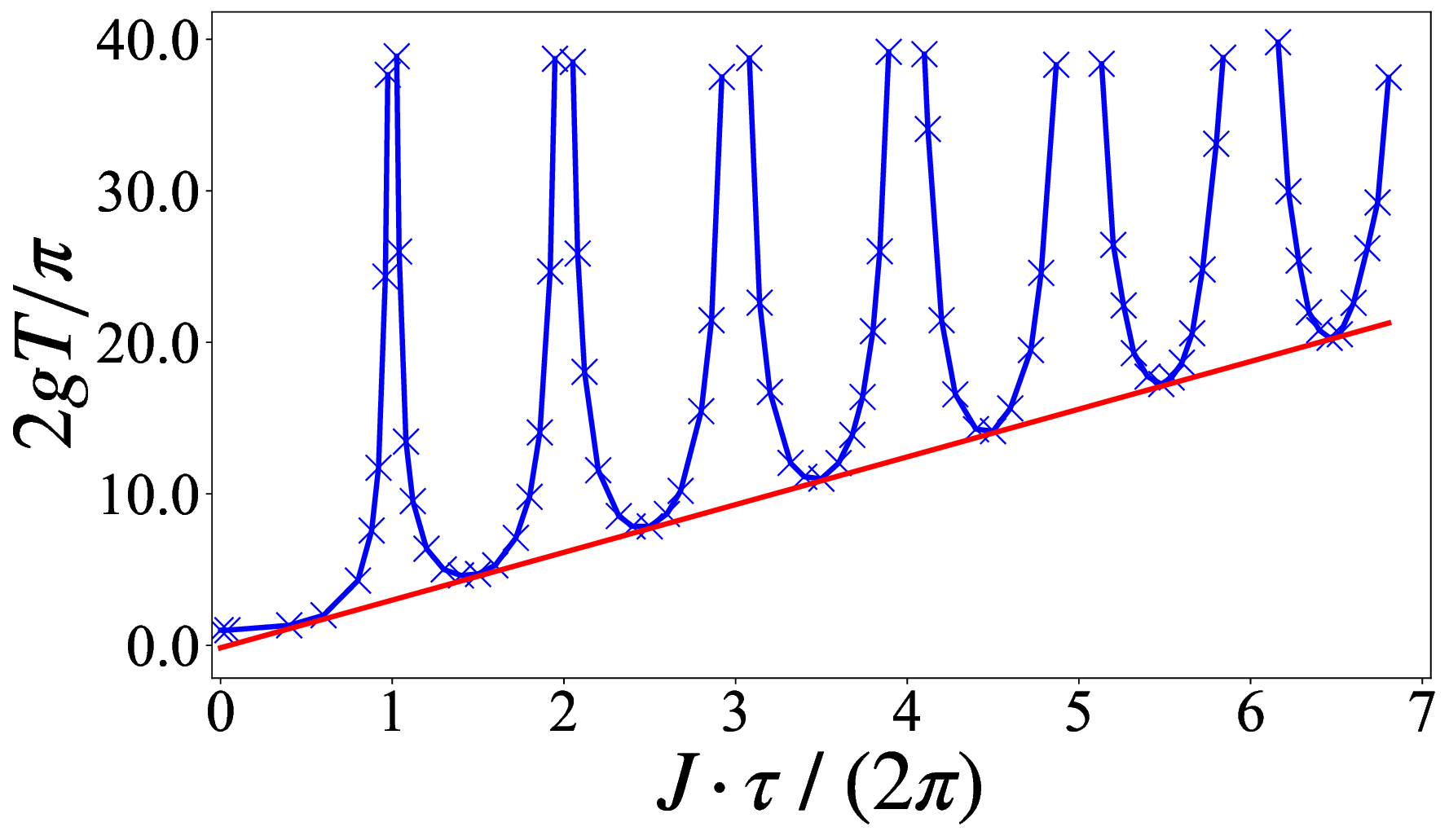}
    \caption{Parameter choice: \(g = 0.01\omega_0\) and $J=0.8\omega_0$. At large scales, the charging time \(T\) increases with the kicking interval \(\tau\), indicating a decrease in the charging speed. A linear fit of \(T\) at the locally optimal points yields \(T \sim 63\tau\). See the main text and Appendix D for further discussion.}
    \label{fig:peaks}
\end{figure}

For arbitrary kicking intervals \(\tau\), we observe numerically that the dynamics of the charger energy \(E_c\) and the battery energy \(E_b\) consistently exhibit sinusoidal behavior (see examples in Appendix C). We fit \(E_c\) using the function
\begin{equation}
    E_c(t) \approx \dfrac{\omega_0}{2}[\cos(\pi t/T) + 1],
\end{equation}
where \(T\) is the time at which the battery first reaches full charge, with \(1/T\) defining the charging speed. By varying the kicking interval \(\tau\), we extract \(T\) from the fitted dynamics and plot it as a function of \(\tau\) in Figure~\ref{fig:peaks}.

At large scales, we observe that \(T\) increases monotonically with \(\tau\). This indicates that the charging speed decreases as the kicks become more widely spaced, while the total energy deposited into the battery remains unchanged at \(\omega_0\). However, the plots also reveal a set of discrete singular peaks.

These singularities occur at discrete kicking intervals \(\tau_n = 2 \pi n / J\), matching the peak positions shown in Figure~\ref{fig:peaks}. More examples are provided in Appendix D.

At these special kicking intervals \(\tau = \tau_n\), if we focus only on the expanded battery, the effective time-evolution operator between two kicks becomes $U(\tau_n) \equiv (\sigma^z_m \otimes \mathbb{I}_b) e^{-i \mathcal{H}_{mb} \tau_n}$.

It can be shown that $U(\tau_n)(|0\rangle \otimes |0\rangle) = i^n|0\rangle \otimes |0\rangle$ and
$U(\tau_n)(|0\rangle \otimes |1\rangle) = |0\rangle \otimes |1\rangle$, meaning that these product states are eigenstates of the evolution operator. Since the modulator remains fixed in the state \(|0\rangle\), the battery states \(|0\rangle\) and \(|1\rangle\) also become eigenstates of the time evolution. This allows one to define a local effective Hamiltonian for the battery, but only at these discrete kicking intervals \(\tau=\tau_n\). However, the energy splitting of this effective Hamiltonian does not match that of the charger, \(\omega_1\), resulting in a finite detuning. Consequently, energy exchange is suppressed, leading to the large values of \(T\) observed at these singular points in Figure~\ref{fig:peaks}. See Appendix D for further discussion.

The valley points in Figure~\ref{fig:peaks} correspond to kicking intervals \(\tau\) that enable relatively fast charging, where the charging time \(T\) exhibits a linear dependence on \(\tau\). A linear fit through these local minima gives \(T/\tau \approx 63\), as shown by the red line. Thus, at large scales, the charging time \(T\) scales linearly with the kicking interval \(\tau\).

\section{Collective Charging and Scalability}

We now extend the modulator-assisted charging protocol from the single-battery setting to a collective many-body architecture. The main question is whether the collective charging advantage of quantum batteries \cite{ferraro2018high} survives under the present modulator-assisted control scheme. In the following, we show that it does.

\begin{figure}[!t]
    \centering
    \includegraphics[width=0.99\linewidth]{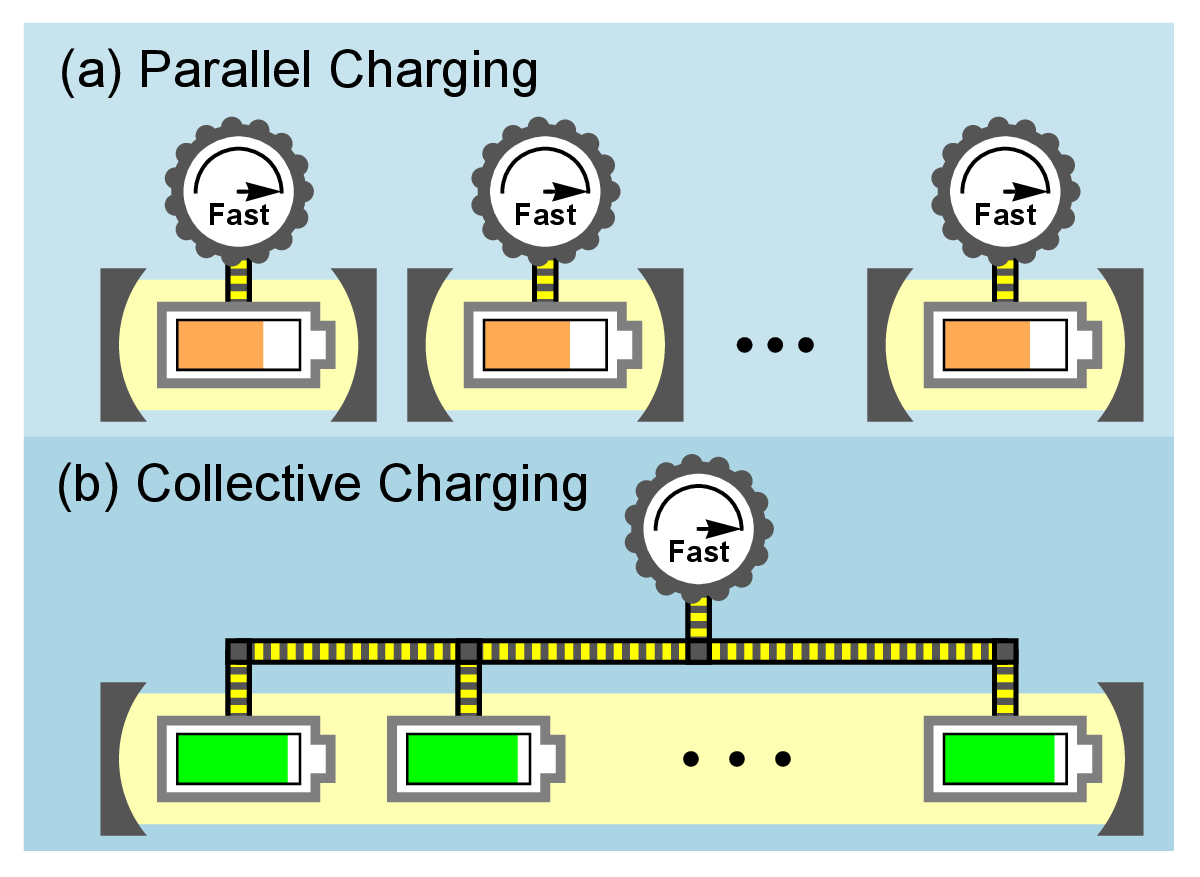}
    \caption{A schematic illustration of our modulator-assisted battery charging protocol when extended to $N$ batteries. \textbf{Panel (a)}: Parallel-charging benchmark. An array of identical modulator-assisted Jaynes-Cummings quantum battery. \textbf{Panel (b)}: Collective charging. A modulator-assisted Tavis-Cummings quantum battery.}
    \label{fig:tcbattery}
\end{figure}

Specifically, we consider a collective architecture in which a single cavity mode acts as a common charger for \(N\) battery qubits, while all battery qubits are coupled to one auxiliary modulator qubit. A schematic illustration of our model is in Figure~\ref{fig:tcbattery}. The total Hamiltonian is
\begin{equation}\label{hmbcN}
    \begin{split}
        \mathcal{H}^{(N)}_{mBc}
        =&-\dfrac{\omega_0}{2}\sigma^z_{m}+\sum_{i=1}^N
        \Big[\dfrac{J}{2}\left(\sigma^x_{m}\sigma^x_{b_i}+\sigma^y_{m}\sigma^y_{b_i}\right)\\
        &-\dfrac{\omega_0}{2}\sigma^z_{b_i}
        +g\left(\sigma^+_{b_i}a_c+\sigma^-_{b_i}a_c^\dagger\right)
        \Big]
        +\omega_1 a_c^\dagger a_c .
    \end{split}
\end{equation}
This corresponds to a modulator-assisted Tavis-Cummings quantum battery. We show that the modulator can provide a global on/off control of the charging process for all \(N\) battery qubits.

We initialize the cavity in the Fock state \(|N\rangle_c\), each battery qubit in its ground state \(|0\rangle_{b_i}\), and the modulator in its ground state \(|0\rangle_m\). As shown in Appendix E, the relevant transition frequencies of the collective modulator-battery sector are \(\omega_0\pm J\sqrt{N}\). When the cavity is tuned to \(\omega_1=\omega_0\), the battery transition is detuned from the charger by \(|\Delta_{bc}(N)|=J\sqrt{N}\). Therefore, for \(J\gg g\), the collective charging process is effectively switched off. This reduces to the single-battery result \(|\Delta_{bc}|=J\) at \(N=1\).

To switch on the charging process, we again apply repeated local unitary kicks \(\sigma^z_m\) to the modulator, interspersed between short intervals of free evolution. As in the single-battery case, these kicks dynamically suppress the modulator-battery exchange interaction. The resulting effective Hamiltonian is
\begin{equation}
    \begin{split}
        \mathcal{H}_{Bc}^{(N)}
        =
        \sum_{i=1}^N
        \Big[
        -\dfrac{\omega_0}{2}\sigma^z_{b_i}
        +g\left(\sigma^+_{b_i}a_c+\sigma^-_{b_i}a_c^\dagger\right)
        \Big]
        +\omega_1 a_c^\dagger a_c.
    \end{split}
\end{equation}
When \(\omega_1=\omega_0\), all battery qubits become resonant with the cavity, and the collective charging process is activated.

A direct simulation of Eq.~\eqref{hmbcN} in the full Hilbert space becomes exponentially costly as \(N\) increases. However, the model has two useful symmetries: (a) permutation symmetry among the \(N\) identical battery qubits and (b) conservation of the total excitation number. By working in the Dicke symmetric basis, Appendix E shows that the dynamics can be restricted to a reduced Hilbert space of dimension \(2N+1\), which enables efficient simulation at larger system sizes.

To quantify the charging performance, we define the maximal stored battery energy and the maximal charging power
\begin{equation}
    E^{(N)}_{\sharp,\mathrm{max}}\equiv\max_t\sum_{i=1}^N E_{b_i}(t),\ P^{(N)}_{\sharp,\mathrm{max}}\equiv\max_t \frac{\sum_i E_{b_i}(t)}{t},
\end{equation}
corresponding to collective charging shown in Figure~\ref{fig:tcbattery}(b). These are compared with the parallel benchmark
\begin{equation}
    E^{(N)}_{\parallel,\mathrm{max}}:=N E^{(1)}_{\sharp,\mathrm{max}},\quad
    P^{(N)}_{\parallel,\mathrm{max}}:=N P^{(1)}_{\sharp,\mathrm{max}},
\end{equation}
corresponding to parallel charging shown in Figure~\ref{fig:tcbattery}(a).

\begin{figure}[!t]
    \centering
    \includegraphics[width=0.95\linewidth]{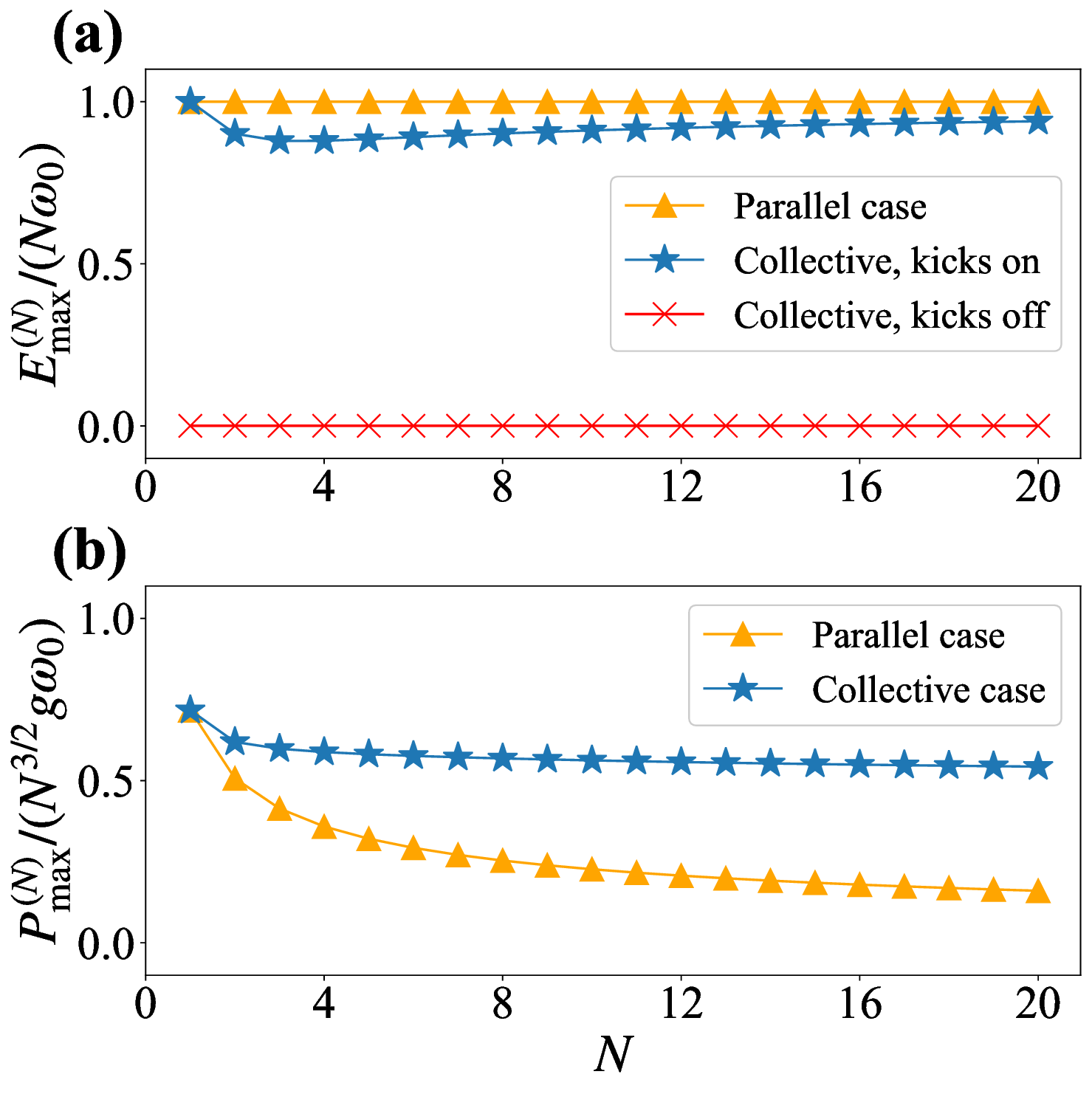}
    \caption{\textbf{Panel (a)}: Maximal stored energy as a function of the battery number \(N\), expressed in units of \(N\omega_0\). The red markers denote the collective charging case without kicks, \(E_{\sharp,\mathrm{max}}^{(N)}(\mathrm{off})\); the blue markers denote the collective charging case with kicks, \(E_{\sharp,\mathrm{max}}^{(N)}(\mathrm{on})\); and the orange markers denote the parallel-charging benchmark, \(E_{\parallel,\mathrm{max}}^{(N)}(\mathrm{on})\). \textbf{Panel (b)}: Maximal charging power as a function of \(N\), expressed in units of \(N^{3/2}g\omega_0\). The blue markers denote the collective charging case, \(P_{\sharp,\mathrm{max}}^{(N)}\), and the orange markers denote the parallel-charging benchmark, \(P_{\parallel,\mathrm{max}}^{(N)}\).}
    \label{fig:scaling}
\end{figure}

The results are in Figure~\ref{fig:scaling}. Figure~\ref{fig:scaling}(a) shows the maximal stored energy in units of \(N\omega_0\) as a function of \(N\). The red markers correspond to the collective architecture without kicks. The maximal stored energy remains zero for all \(N\), showing that the charging interaction is effectively suppressed in the absence of kicks. In contrast, the blue markers correspond to the kicked collective charging case, and the orange markers correspond to the parallel benchmark. In both cases, the rescaled stored energy saturates to a constant, indicating linear scaling with system size, \(E^{(N)}_{\sharp,\mathrm{max}}\propto N\) and \(E^{(N)}_{\parallel,\mathrm{max}}\propto N\).

Figure~\ref{fig:scaling}(b) shows the maximal charging power in units of \(N^{3/2}g\omega_0\). In the collective charging case (blue markers), the rescaled power approaches a constant, indicating the scaling \(P^{(N)}_{\sharp,\mathrm{max}}\propto N^{3/2}\). In contrast, the parallel benchmark (orange markers) scales as \(P^{(N)}_{\parallel,\mathrm{max}}\propto N\). This is consistent with the expected collective charging advantage \cite{ferraro2018high}.

Taken together, Figures~\ref{fig:scaling}(a) and \ref{fig:scaling}(b) show that the modulator-assisted protocol remains effective in the many-body setting: without kicks, the charging process is strongly suppressed; with kicks, collective charging is restored and retains its \(N^{3/2}\) power scaling. The collective charging advantage is thus reflected in the time required to reach maximal battery energy, which scales as \(1/\sqrt{N}\).

\section{Experimental Feasibility in an NV-\texorpdfstring{$\Cs$}{13C} Platform}

Our proposal can be naturally connected to a solid-state spin platform based on an NV center coupled to surrounding $\Cs$ nuclear spins~\cite{bradley2019ten}. In this setting, the NV electronic spin provides the modulator degree of freedom, while selected nearby $\Cs$ nuclei can be divided into two subsets: a battery subset and a charger subset. The battery subset plays the role of the many-battery qubits, whereas the charger subset is treated collectively and serves as an effective common charger mode. This architecture provides a realistic route for scaling the stored energy with the number of battery units, addressing the many-body regime considered in our model.

At the microscopic level, the NV center and surrounding $\Cs$ nuclei are described by the standard hyperfine Hamiltonian, in which the NV electronic spin couples individually to each nuclear spin. By restricting the NV center to a two-level subspace and applying suitable microwave and radio-frequency control, one can engineer selective effective couplings between the modulator and chosen nuclear spins, for instance via spin-locking and dynamical decoupling techniques~\cite{ajoy2015atomic}. In particular, moving to the rotating frame leads to the desired modulator-battery interaction,
\begin{equation}
    \mH_{mB}^{\mathrm{eff}}
    \approx
    -\dfrac{\omega_0}{2}\sigma_m^z
    -
    \sum_{i=1}^N
    \dfrac{\omega_0}{2}\sigma_{b_i}^z
    +
    \dfrac{J}{2}
    \sum_{i=1}^N
    \left(
        \sigma_m^x\sigma_{b_i}^x+\sigma_m^y\sigma_{b_i}^y
    \right).
\end{equation}
In addition, geometric placement of nuclear spins can be used to suppress undesired hyperfine components, for example via near-magic-angle configurations~\cite{dreau2012high}.

A direct battery-charger flip-flop interaction is not natively present, but it can be generated perturbatively through control engineering. Specifically, by using the NV center as a mediator and applying a short Floquet commutator sequence, one can induce an effective exchange interaction between the battery nuclei and the charger nuclei. In the low-excitation regime, the charger subset can be reduced to a collective bright mode \(b\), yielding an effective charger sector of the form
\begin{equation}
    \mH_{BC}^{\mathrm{eff}}
    \approx
    g\sum_{i=1}^N
    \left(
        \sigma_{b_i}^+ a_c+\sigma_{b_i}^- a_c^\dagger
    \right)
    +
    \omega_1 a_c^\dagger a_c.
\end{equation}
In this representation, the charger acts as a shared energy reservoir whose effective capacity increases with the size of the nuclear ensemble, enabling scalable energy transfer to multiple battery units.

As a result, the full many-battery model can be reconstructed in the sense of effective Floquet simulation by combining the modulator-battery and battery-charger sectors stroboscopically. The repeated local kicks applied to the modulator in our protocol can be implemented directly as short resonant microwave pulses on the NV electronic spin, thereby realizing the required \(\sigma_m^z\)-type control between successive effective evolution steps.

State initialization in this platform can be achieved through optical polarization of the NV center together with nuclear spin polarization mechanisms such as the excited-state level anti-crossing (ESLAC)~\cite{jacques2009dynamic}. In this way, the modulator can be prepared in a definite computational state, the battery nuclei can be initialized close to their ground states, and the charger subset can be prepared in a controlled low-excitation collective state.

In a realistic setting, decoherence and losses will inevitably affect the dynamics. In particular, relaxation and dephasing of the NV electronic spin, decoherence of the nuclear spins, and loss of excitations from the collective charger mode will reduce the achievable stored energy and charging power. The relevant charging dynamics is expected to remain observable as long as the system parameters \(J\), \(g\), and \(\omega_0\) are sufficiently larger than the corresponding noise rates. Consequently, noise primarily leads to quantitative reductions in performance, rather than a breakdown of the effective-Hamiltonian description.

Taken together, these ingredients show that the proposed modulator-assisted many-battery charging architecture is compatible with an experimentally motivated NV-$\Cs$ platform at the level of effective Hamiltonian engineering. This makes the scheme not only conceptually scalable, but also potentially relevant to near-term solid-state spin implementations. A detailed derivation of the effective interactions, repeated-kick implementation, initialization procedure, and noise modeling is provided in Appendix~F.

\section{Discussion and Summary}

We first note that the use of repeated control operations in quantum batteries has been explored in a different context, where the quantum Zeno effect is employed during the storing phase to protect the stored energy \cite{gherardini2020stabilizing}, in close analogy to refocusing techniques such as spin echo in NMR systems \cite{hahn1950spin,levitt1979nmr}. 

In contrast, the present protocol exploits Zeno-like control during the charging phase rather than the storing phase. This distinction has two important consequences. First, the total number of required control operations is significantly reduced, since the charging duration is typically much shorter than the storage time. Second, the role of the control is not merely to suppress dynamics, but to actively reshape the effective Hamiltonian governing energy transfer. In this sense, the modulator serves as an indirect control interface that converts local operations into nonlocal regulation of the charger-battery coupling.

Another key feature of our approach is that it does not rely on the idealized limit of infinitely fast control. While the dense-kicking regime provides a clear effective-Hamiltonian picture, our analysis shows that the protocol remains operational for finite kicking intervals. Increasing the interval $\tau$ leads to a gradual reduction in charging speed, but does not destroy the ability to switch the energy-transfer channel on and off. This robustness relaxes the requirement of ultra-fast control and broadens the range of experimentally accessible regimes.

From a broader perspective, the protocol introduces a distinct control paradigm for quantum batteries: instead of directly tuning the charger, the battery, or their interaction, one achieves control through an auxiliary degree of freedom that is locally accessible. This is particularly advantageous in architectures where direct manipulation of the energy-storage subsystem is difficult, costly, or introduces additional noise. Because the charger-battery interaction remains always on, the scheme also avoids the need for time-dependent coupling elements, which can be a major experimental constraint in many platforms.

Importantly, this indirect-control mechanism is compatible with collective many-body architectures. As shown in this work, the protocol extends naturally to a modulator-assisted Tavis-Cummings battery, where a single modulator controls the energy flow between a common charger and an array of $N$ battery qubits. In this setting, the protocol preserves the characteristic collective enhancement of charging power, with $P_{\sharp}^{(N)} \propto N^{3/2}$ \cite{ferraro2018high,gemme2023off,hogan2024quench,yang2024three,seidov2024quantum}, while still enabling global on/off switching through local operations.

In summary, we have introduced a modulator-assisted control framework for charger-mediated quantum batteries, in which repeated local unitary operations on an auxiliary qubit induce a Zeno-like reshaping of the effective Hamiltonian. This enables dynamic switching of energy transfer without direct intervention on the charger-battery subsystem. The protocol operates beyond the idealized fast-control limit, is scalable to collective many-body settings, and is compatible with experimentally motivated platforms. In particular, our analysis suggests that the key ingredients of the protocol can be engineered in solid-state spin systems such as an NV-\(\Cs\) platform through effective-Hamiltonian control. More broadly, our results suggest a general strategy for controlling interaction-driven processes in composite quantum systems through indirect, local, and always-on control architectures.

\section*{Author Contributions}
Conceptualization: S.X.; Methodology: S.X.~and M.S.; Formal analysis: S.X., A.A.~and S.K.; Investigation: S.X.~and M.S.; Visualization: S.X.; Writing---original draft: S.X.~and A.A.; Writing---review \& editing: M.S.~and S.K.; Funding acquisition: S.K.

\section*{Acknowledgments}
S.X.~thanks Peter W.~Milonni and Yuan-Yuan Zhao for valuable discussions. S.K.~acknowledges support from the US Department of Energy, Office of Basic Energy Sciences, through the Quantum Photonic Integrated Design Center (QuPIDC) EFRC award DE-SC0025620.

\bibliographystyle{quantum}
\bibliography{battery}

\onecolumn\newpage
\appendix

\section{Derivation of Eq.~(5)}
\renewcommand{\theequation}{A\arabic{equation}}
\setcounter{equation}{0}
\renewcommand{\thefigure}{A\arabic{figure}}
\setcounter{figure}{0}

The interaction Hamiltonian $\mathcal{V}_{bc}=g(\sigma^+_b\sigma^-_c+\sigma^-_b\sigma^+_c)$ can be expressed more explicitly as:
\begin{equation}\label{inter}
    g(\mathbb{I}_m\otimes|1\rangle \langle0|_b\otimes|0\rangle\langle1|_c+\text{H.c.}),
\end{equation}
where H.c.~is the Hermitian conjugate.

The goal is to decompose the vectors of the joint Hilbert space $mb$ in the eigenbasis of $\mathcal{H}_{mb}$ in Eq.~(1). The eigenvectors of $\mathcal{H}_{mb}$ are:
\begin{equation}
\begin{aligned}
|v_0\rangle &= |0\rangle_m|0\rangle_b, 
&\ \text{with }\lambda_0 &= -\omega_0,\\
|v_1\rangle &= \frac{1}{\sqrt{2}}\bigl(|0\rangle_m|1\rangle_b-|1\rangle_m|0\rangle_b\bigr), 
&\ \text{with }\lambda_1 &= -J,\\
|v_2\rangle &= \frac{1}{\sqrt{2}}\bigl(|0\rangle_m|1\rangle_b+|1\rangle_m|0\rangle_b\bigr), 
&\ \text{with }\lambda_2 &= J,\\
|v_3\rangle &= |1\rangle_m|1\rangle_b, 
&\ \text{with }\lambda_3 &= \omega_0.
\end{aligned}
\end{equation}

To achieve this, we first observe that 
\begin{equation}
    \begin{split}
        &|01\rangle_{mb}=\dfrac{1}{\sqrt{2}}\bigl(|v_2\rangle_{mb}+|v_1\rangle_{mb}\bigl),\\
        &|10\rangle_{mb}=\dfrac{1}{\sqrt{2}}\bigl(|v_2\rangle_{mb}-|v_1\rangle_{mb}\bigl).\\
    \end{split}
\end{equation}

This decomposes the operator $\mathbb{I}_m\otimes|1\rangle\langle0|_b$ into the following four terms:
\begin{equation}
    \begin{split}
        &\mathbb{I}_m\otimes|1\rangle\langle0|_b\\
        =&|0\rangle\langle0|_m\otimes|1\rangle\langle0|_b+|1\rangle\langle1|_m\otimes|1\rangle\langle0|_b\\
        =&\dfrac{1}{\sqrt{2}}\bigl(|v_1\rangle\langle v_0|+|v_2\rangle\langle v_0|+|v_3\rangle\langle v_2|-|v_3\rangle\langle v_1|\bigl).
    \end{split}
\end{equation}
Together, the interaction Hamiltonian becomes
\begin{equation}\label{12terms}
    \begin{aligned}
        \mathcal{V}_{bc}
        = \frac{g}{\sqrt{2}} \Big[ \,
        &\big(
        |v_1\rangle\langle v_0|_{mb}
        + |v_2\rangle\langle v_0|_{mb}
        + |v_3\rangle\langle v_2|_{mb} \\
        &\quad - |v_3\rangle\langle v_1|_{mb}
        \big)\otimes |0\rangle\langle 1|_c
        + \mathrm{H.c.}
        \Big].
    \end{aligned}
\end{equation}
This is precisely Eq.~(5).


\section{Derivation of Eq.~(7)}
\renewcommand{\theequation}{B\arabic{equation}}
\setcounter{equation}{0}
\renewcommand{\thefigure}{B\arabic{figure}}
\setcounter{figure}{0}

Consider the unitary kicks are applied to the modulator at every time of $\tau$. For a duration of $2\tau$, the total time evolution operator is
\begin{equation}\label{u2tau}
    \begin{split}
        U(2\tau)=&\sigma^z_me^{-i\mH_{mbc}\tau}\sigma^z_me^{-i\mH_{mbc}\tau}\\
        =&e^{-i\sigma^z_m\mH_{mbc}\sigma^z_m\tau}e^{-i\mH_{mbc}\tau},
    \end{split}
\end{equation}
where the last line is due to the unitarity of $\sigma^z_m$.

One observes the relations:
\begin{equation}\label{xyz}
    \begin{split}
        &\sigma^z_m\sigma^x_m\sigma^z_m=-\sigma^x_m,\\
        &\sigma^z_m\sigma^y_m\sigma^z_m=-\sigma^y_m,\\
        &\sigma^z_m\sigma^z_m\sigma^z_m=+\sigma^z_m.
    \end{split}
\end{equation}

Substituting Eq.~\eqref{xyz} to Eq.~\eqref{u2tau}, one obtains:
\begin{equation}
    \begin{split}
        U(2\tau)=&e^{-i\mH'_{mbc}\tau}e^{-i\mH_{mbc}\tau},\\
        \text{with}\ H_{mbc} =& -\dfrac{\omega_0}{2} \sigma_m^z + \dfrac{J}{2} \left(\sigma_m^x \sigma_b^x + \sigma_m^y \sigma_b^y\right) - \dfrac{\omega_0}{2} \sigma_b^z\\ 
        &+ \dfrac{g}{2}(\sigma^{x}_b\sigma^{x}_c+\sigma^{y}_b\sigma^{y}_c) - \dfrac{\omega_1}{2}\sigma^z_c,\\
        \text{and}\ H'_{mbc} =& -\dfrac{\omega_0}{2} \sigma_m^z - \dfrac{J}{2} \left(\sigma_m^x \sigma_b^x + \sigma_m^y \sigma_b^y\right) - \dfrac{\omega_0}{2} \sigma_b^z\\ 
        &+ \dfrac{g}{2}(\sigma^{x}_b\sigma^{x}_c+\sigma^{y}_b\sigma^{y}_c) - \dfrac{\omega_1}{2}\sigma^z_c.
    \end{split}
\end{equation}

The leading order in the Baker-Campbell-Hausdorff (BCH) expansion gives
\begin{equation}
    \begin{split}
        U(2\tau)=&e^{-i\mH'_{mbc}\tau}e^{-i\mH_{mbc}\tau}\equiv e^{-imH_{mbc}^\text{eff}\tau},\text{ with}\\
        H_{mbc}^\text{eff}=&\dfrac{1}{2}\bigl(\mH'_{mbc}+\mH_{mbc}\bigl)+\mathcal{O}(\tau)\\
        \approx&-\frac{\omega_0}{2}\sigma_m^z-\frac{\omega_0}{2}\sigma_b^z+\dfrac{g}{2}(\sigma^x_b\sigma^x_c+\sigma^y_b\sigma^y_c)-\frac{\omega_1}{2}\sigma_c^z.
    \end{split}
\end{equation}
This is precisely Eq.~(7).


\section{Numerical fitting of the battery's energy dynamics}
\renewcommand{\theequation}{C\arabic{equation}}
\setcounter{equation}{0}
\renewcommand{\thefigure}{C\arabic{figure}}
\setcounter{figure}{0}

When the kicks on the modulator are more widely spaced---i.e., when the kicking interval $\tau$ increases---the charging speed becomes slower, while the charging capacity remains unchanged.

Both the charger's energy $E_c$ and the battery's energy $E_b$ exhibit sinusoidal dynamics. Furthermore, the two energies are symmetric: their sum remains constant over time. This implies that as one increases, the other decreases.

Hence, it suffices to consider only the charger's energy, which we approximate using the fit function
\begin{equation}
    E_c(t) \approx \frac{\omega_0}{2} \left[ \cos\left( \pi t/T \right) + 1 \right],
\end{equation}
where $T$ denotes the first time when the battery becomes maximally charged with energy $\omega_0$. 

\begin{figure}[!b]
    \centering
    \includegraphics[width=0.4\linewidth]{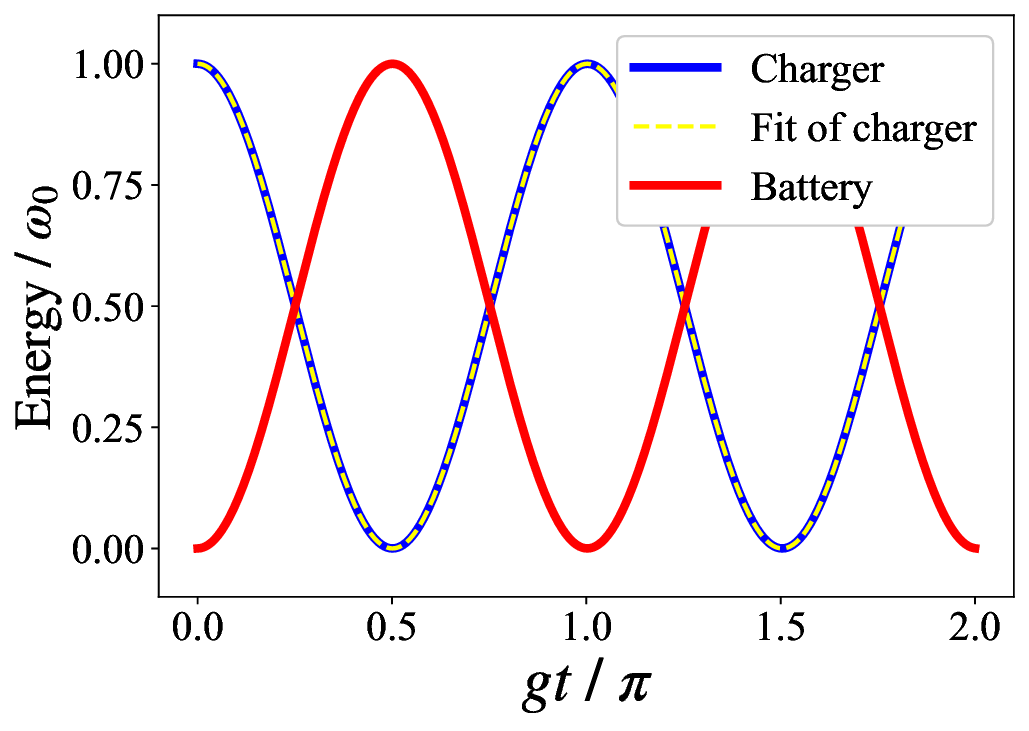}
    \includegraphics[width=0.4\linewidth]{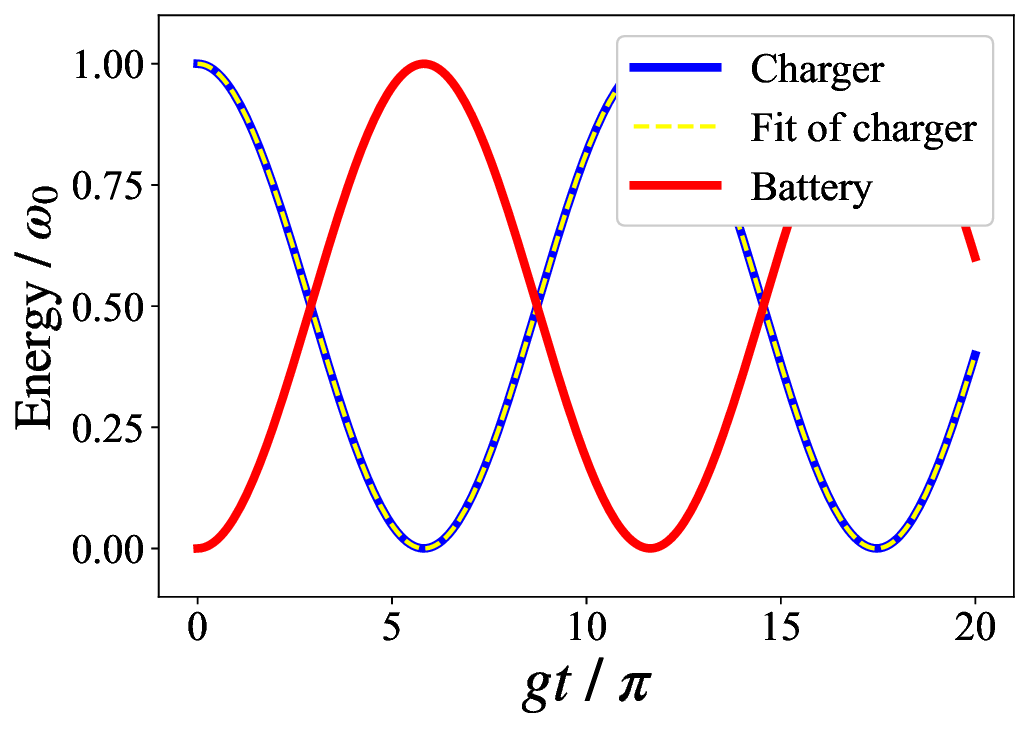}
    \caption{\textbf{Left}: Setting $\omega_1=\omega_0$, $g=0.01\omega_0$. The kicking interval is $\tau=\pi/(1000g)$. The battery gets maximally charged at $T=0.5\pi/g$. \textbf{Right}: Setting $\omega_1=\omega_0$, $g=0.01\omega_0$. The kicking interval is $\tau=84\pi/(1000g)$. The battery gets maximally charged at $T=5.82\pi/g$.}
    \label{fig:sin12}
\end{figure}

One may check the examples for different $\tau$ values in Figure~\eqref{fig:sin12}.

\section{Analysis of the Singular Peaks in Figure~3}
\renewcommand{\theequation}{D\arabic{equation}}
\setcounter{equation}{0}
\renewcommand{\thefigure}{D\arabic{figure}}
\setcounter{figure}{0}

The eigenbasis of $\mathcal{H}_{mb}$ is given by:
\begin{equation}
\begin{aligned}
|v_0\rangle &= |0\rangle_m|0\rangle_b, 
&\ \text{with }\lambda_0 &= -\omega_0,\\
|v_1\rangle &= \frac{1}{\sqrt{2}}\bigl(|0\rangle_m|1\rangle_b-|1\rangle_m|0\rangle_b\bigr), 
&\ \text{with }\lambda_1 &= -J,\\
|v_2\rangle &= \frac{1}{\sqrt{2}}\bigl(|0\rangle_m|1\rangle_b+|1\rangle_m|0\rangle_b\bigr), 
&\ \text{with }\lambda_2 &= J,\\
|v_3\rangle &= |1\rangle_m|1\rangle_b, 
&\ \text{with }\lambda_3 &= \omega_0.
\end{aligned}
\end{equation}

\medskip
\noindent
\textbf{Main conclusion.} The kicking intervals $\tau$ that lead to the singular peaks, as can be observed in Figure~3, are denoted by $\tau_n$. The values of $\tau_n$ are determined only by the eigenvalues $\lambda_1$ and $\lambda_2$, not by $\lambda_0$ or $\lambda_3$. These quantized values occur at
\begin{equation}
    \tau_n = \frac{2\pi n}{\lambda_2}\equiv \frac{2\pi n}{J}, \quad n \in \mathbb{Z}^+,
\end{equation}
where $n$ is a positive integer.

Here we analyze the origin of the singular peaks by dividing the values of $\tau$ into several regimes (the analysis is based on $\omega_1=\omega_0$):

\begin{enumerate}[label={(\arabic*)}]
\item {\bf Fast-kicking regime, $\tau\rightarrow0$:}

In the fast-kicking limit, it has been shown that the local effective Hamiltonian for the battery becomes $-\frac{\omega_0}{2}\sigma^z_b$, with the battery initialized in the ground state $|0\rangle_b$. Meanwhile, the charger's Hamiltonian is $-\frac{\omega_0}{2} \sigma^z_c$, with the charger initialized in the excited state $|1\rangle_b$. Under the interaction Hamiltonian Eq.~(8), energy exchange between the charger and the battery is enabled in this regime.

    \item {\bf No-kicking regime, $\tau\rightarrow\infty$:}

In the absence of kicking, the possible transitions for the battery, as given in Eq.~\eqref{12terms}, namely
$|v_0\rangle \leftrightarrow |v_1\rangle$, 
$|v_0\rangle \leftrightarrow |v_2\rangle$, 
$|v_1\rangle \leftrightarrow |v_3\rangle$, and 
$|v_2\rangle \leftrightarrow |v_3\rangle$, all have energy differences that do not match that of the charger transition $|0\rangle \leftrightarrow |1\rangle$. As a result, energy exchange is suppressed due to detuning.

\item \textbf{Intermediate regime: for general values of $0 < \tau < \infty$}

For a general kicking interval $\tau$, the total evolution operator acting on the modulator-battery system consists of two parts: (a) a free evolution under the Hamiltonian $H_{mb}$ for a duration $\tau$, and (b) a control kick represented by  $\sigma^z_m$ applied to the modulator. The combined evolution operator is given by:
\begin{equation}
    U(\tau) \equiv \sigma^z_m\cdot e^{-i\mH_{mb}\tau}.
\end{equation}

Since the modulator is initialized in the state $|0\rangle_m$, the action of the kick effectively preserves its state. Consequently, we only need to consider the two states $|0\rangle_m|0\rangle_b$ and $|0\rangle_m|1\rangle_b$ in the analysis.

It is straightforward to verify that the state $|0\rangle_m|0\rangle_b$ is an eigenvector of $U(\tau)$ with eigenvalue $-e^{-i\omega_0\tau}$.

On the other hand, the state $|0\rangle_m|1\rangle_b$ is generally \emph{not} an eigenvector of $U(\tau)$. Therefore, in this intermediate regime, the local effective Hamiltonian for the battery is generally ill-defined. The resulting dynamics involve complex coupling between the modulator and the charger, making the energy exchange process complex and lacking a simple physical interpretation.

\item \textbf{Intermediate regime: for $\tau= \tau_n\equiv 2\pi n / J$, with $n \in \mathbb{Z}^+$}

At these discrete values of $\tau$, it follows that 
\begin{equation}
    \begin{split}
        U(\tau_n)|0\rangle_m|0\rangle_b=&-e^{-i\omega_0\tau}|0\rangle_m|0\rangle_b,\\
        U(\tau_n)|0\rangle_m|1\rangle_b =& |0\rangle_m|1\rangle_b.
    \end{split}
\end{equation}

\begin{figure}[!t]
    \centering
    \includegraphics[width=0.5\linewidth]{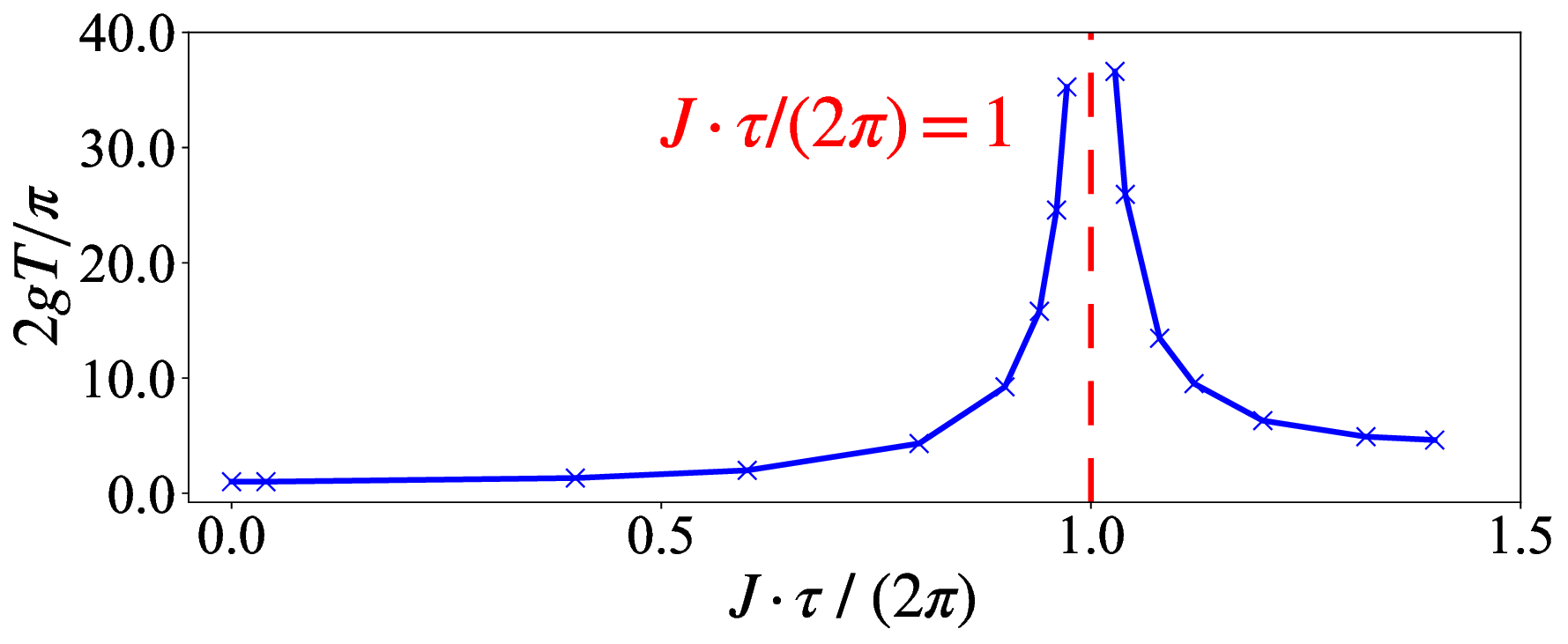}
    \includegraphics[width=0.5\linewidth]{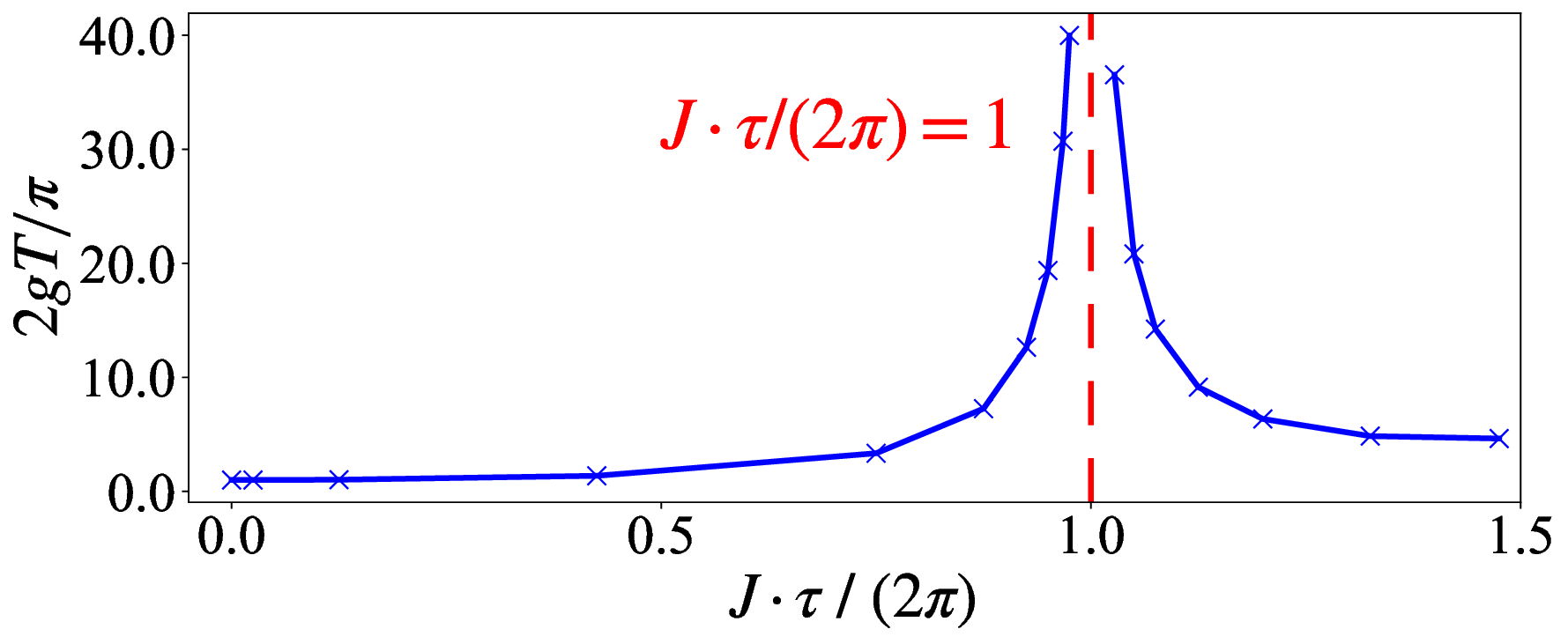}
    \caption{\textbf{Top}: We choose $\omega_0 = 1.5$ and $J = 0.8$, while fixing the system parameters $\omega_1=\omega_0$ and $g = 0.01$. Under these conditions, it can be shown that the first singular peak occurs at $J\tau/(2\pi)=1$. \textbf{Bottom}: We choose $\omega_0 = 1.0$ and $J = 0.5$, while fixing the system parameters $\omega_1=\omega_0$ and $g = 0.01$. Under these conditions, it can be shown that the first singular peak occurs at $J\tau/(2\pi)=1$. These results match the analysis.}
    \label{fig:peaks12}
\end{figure}

Therefore, at these specific values of $\tau$, both $|0\rangle_m|0\rangle_b$ and $|0\rangle_m|1\rangle_b$ are eigenstates of the effective evolution operator $U(\tau_n)$. This implies that, under the assumption that the modulator remains frozen in the state $|0\rangle_m$, a well-defined local effective Hamiltonian exists for the charger, with eigenstates $|0\rangle$ and $|1\rangle$.

However, the energy splitting of this effective charger Hamiltonian does not match that of the battery, which is $\omega_1$. The resulting detuning prevents resonant energy exchange between the charger and the battery. This mismatch explains the appearance of singularities in Figure~3.
\end{enumerate}

To further validate this analysis, we vary the values of $\omega_0$ and $J$ while fixing $\omega_1=\omega_0$ and $g$, and observe that the first singular point consistently appears at $\tau = 2\pi / J$. See Figure~\ref{fig:peaks12}.


\section{Many-battery Hamiltonian with symmetry}
\renewcommand{\theequation}{E\arabic{equation}}
\setcounter{equation}{0}
\renewcommand{\thefigure}{E\arabic{figure}}
\setcounter{figure}{0}
\setcounter{subsection}{0}

We consider $N$ batteries (labeled $b_1,\cdots b_N$) coupled to one single cavity (labeled $c$) that serves as a charger. The $N$ batteries are coupled to one single modulator (labeled $m$). The total Hamiltonian is given by
\begin{equation}\label{appHmbcN}
    \begin{split}
        \mathcal{H}^{(N)}_{mBc}
        =&-\frac{\omega_0}{2}\sigma^z_{m}+\sum_{i=1}^N
        \Big[\frac{J}{2}\left(\sigma^x_{m}\sigma^x_{b_i}+\sigma^y_{m}\sigma^y_{b_i}\right)\\
        &-\frac{\omega_0}{2}\sigma^z_{b_i}
        +g\left(\sigma^+_{b_i}a_c+\sigma^-_{b_i}a_c^\dagger\right)
        \Big]
        +\omega_1 a_c^\dagger a_c .
    \end{split}
\end{equation}

We initialize the cavity in the Fock state \(|N\rangle_c\), each battery qubit in its ground state \(|0\rangle_{b_i}\), and the modulator in its ground state $|0\rangle_m$. The initial state is
\begin{equation}
    |0\rangle_{m}\otimes\prod_{i=1}^N|0\rangle_{b_i}\otimes|N\rangle_c.
\end{equation}

\subsection{Simplifying simulation using Dicke's symmetry}
A direct simulation of Eq.~\eqref{appHmbcN} is exponentially costly in the full Hilbert space. Indeed, the modulator and the $N$ battery qubits contribute a factor $2^{N+1}$, and a truncated cavity Hilbert space contributes an additional factor $N$ from the photon-number basis, leading to an scaling $\sim\mathcal{O}(N2^{N})$. However, the present model possesses two key symmetries that allow for a dramatic reduction of the effective Hilbert space:
\begin{itemize}
    \item permutation symmetry among the \(N\) identical modulator-battery units, and
    \item conservation of the total excitation number.
\end{itemize}

The first is permutation symmetry among the $N$ identical modulator-battery units. We define the collective operators
\begin{equation}
    S^+_B=\sum_{i=1}^N\sigma^+_{b_i},\ S^-_B=\sum_{i=1}^N\sigma^-_{b_i},\ S^z_B=\sum_{i=1}^N\sigma^z_{b_i}.
\end{equation}
The Hamiltonian Eq.~\eqref{appHmbcN} becomes
\begin{equation}\label{appHmbcN2}
    \begin{split}
        \mathcal{H}^{(N)}_{mBc}=&-\dfrac{\omega_0}{2}\sigma^z_m+J(\sigma^+_mS^-_B+\sigma^-_mS^+_B)\\
        &+g(S^+_Ba_c+S^-_Ba_c^\dagger)+\omega_1a^\dagger_ca_c.
    \end{split}
\end{equation}

The second is conservation of the total excitation number. Since both the modulator-battery coupling and the battery-cavity coupling are of Jaynes-Cummings type, the total excitation number remains fixed throughout the evolution, given by
\begin{equation}
    n_m+n_B+n_c=N .
\end{equation}
Here, $n_m=\{0,1\}$ indicates that the modulator is in the state $|0\rangle_m$ or $|1\rangle_m$. $n_B$ is the total excitation number for the symmetric battery state. $n_c$ is the photon number in the cavity. They add up to the initial photon number $N$.

Therefore, a general symmetric basis state can be written as $|n_m,n_B\rangle$. In particular, the ground state of the global modulator-battery system corresponds to $|0,0\rangle$.

This symmetry-adapted construction reduces the Hilbert-space dimension $D_N$ from exponential to linear scaling, $D_N=2N+1$. Specifically, 
\begin{itemize}
    \item When $n_m=0$, $n_B$ can take the following values
    \begin{equation}
        n_B=0,1,\cdots,N.
    \end{equation}
    \item When $n_m=1$, $n_B$ can take the following values
    \begin{equation}
        n_B=0,1,\cdots,N-1.
    \end{equation}
\end{itemize}
Therefore, the reduced symmetric Hilbert space is $2N+1$ dimensional.

The action of Eq.~\eqref{appHmbcN} or \eqref{appHmbcN2} on a general reduced basis state $|n_m,n_B\rangle$ is given by
\begin{equation}
    \begin{split}
    &\mathcal{H}^{(N)}_{mBc}\ket{n_m,n_B}\\
    =&\left[\omega_0\left(n_a-\frac{1}{2}\right)
    +\omega_0\left(n_B-\frac{N}{2}\right)
    +\omega_1 n_c\right]\ket{n_m,n_B}
    \nonumber\\
    &+J\sqrt{n_B\bigl(N-n_B+1\bigr)}\,
    \ket{n_m+1,n_B-1}
    \nonumber\\
    &+J\sqrt{(N-n_B)(n_B+1)}\,
    \ket{n_m-1,n_B+1}
    \nonumber\\
    &+g\sqrt{(N-n_B)(n_B+1)\,n_c}\,
    \ket{n_m,n_B+1}
    \nonumber\\
    &+g\sqrt{n_B\bigl(N-n_B+1\bigr)(n_c+1)}\,
    \ket{n_m,n_B-1}.
    \end{split}
\end{equation}
Specifically, $|n_m,n_B\rangle\equiv0$ whenever $(n_m,n_B)$ lies outside the allowed values.

Now we consider the unitary kicks applied to the modulators. We define the kick operator to be $\sigma^z_{m}$. The action of this operator on a general basis state $|n_m,n_B\rangle$ is given by
\begin{equation}
    \begin{split}
        \sigma^z_m|n_m,n_B\rangle=(-1)^{n_m}|n_m,n_B\rangle.
    \end{split}
\end{equation}

The above techniques together enable efficient calculations of our modulator-assisted charging paradigm to a large number of batteries.

\subsection{Collective normal-mode splitting in the single-excitation sector}

A key feature of the many-battery Hamiltonian is the emergence of a collective enhancement of the effective coupling between the modulator and the symmetric battery mode. This can be seen clearly in the single-excitation sector.

Starting from the global ground state $|0,0\rangle$, the single-excitation subspace is spanned by two symmetric states
\begin{equation}
    |n_m=1,n_B=0\rangle, \quad |n_m=0,n_B=1\rangle,
\end{equation}
which correspond to one excitation in the modulator and one collective excitation in the batteries, respectively.

Restricting Eq.~\eqref{appHmbcN2} to this subspace, the Hamiltonian takes the form
\begin{equation}
    H^{(1)} =
    \omega_0
    \begin{pmatrix}
        1 & 0 \\
        0 & 1
    \end{pmatrix}
    +
    J\sqrt{N}
    \begin{pmatrix}
        0 & 1 \\
        1 & 0
    \end{pmatrix},
\end{equation}
where the factor $\sqrt{N}$ arises from the collective matrix element
$\langle 0,1|S_B^+|0,0\rangle = \sqrt{N}$.

Diagonalizing this $2\times2$ Hamiltonian yields two collective eigenmodes with energies
\begin{equation}
    E_{\pm} = \omega_0 \pm J\sqrt{N}.
\end{equation}
Therefore, the transition frequencies from the ground state $|0,0\rangle$ to the single-excitation eigenstates are
\begin{equation}
    \omega_{\pm} = \omega_0 \pm J\sqrt{N}.
\end{equation}

When this collective modulator-battery is coupled to a charger with frequency $\omega_0$, the detuning is thus
\begin{equation}
    |\Delta_{bc}(N)|=J\sqrt{N}.
\end{equation}
This demonstrates an $\sqrt{N}$-enhancement in the battery-charger detuning in the absence of external kicks, which comes from the collective coupling between the modulator and the symmetric Dicke mode of the batteries.


\section{Potential experimental realization}
\renewcommand{\theequation}{E\arabic{equation}}
\setcounter{equation}{0}
\renewcommand{\thefigure}{E\arabic{figure}}
\setcounter{figure}{0}
\setcounter{subsection}{0}

\subsection{Native Physical Hamiltonian}
We consider a hybrid spin system consisting of a single NV center electronic spin and several surrounding $\Cs$ nuclear spins. The microscopic Hamiltonian is
\begin{equation}
    \mH
    =
    D S_z^2
    - \gamma_e B_z S_z
    - \sum_{\mu} \gamma_n B_z I_{\mu z}
    + \sum_{\mu} \mathbf{S}\cdot A_\mu \cdot \mathbf{I}_\mu .
\end{equation}
Here, \(D \approx 2.87~\mathrm{GHz}\) is the zero-field splitting of the NV center electronic spin, \(B_z\) is an external static magnetic field applied along the NV symmetry axis, and \(\gamma_e\) and \(\gamma_n\) are the gyromagnetic ratios of the NV electronic spin and the $\Cs$ nuclear spins, respectively. The index \(\mu\) labels the individual $\Cs$ nuclei surrounding the NV center. Moreover, \(\mathbf{S}=(S_x,S_y,S_z)\) denotes the spin-1 operator of the NV center, \(\mathbf{I}_\mu=(I_{\mu x},I_{\mu y},I_{\mu z})\) denotes the spin operator of the \(\mu\)-th $\Cs$ nucleus, and \(A_\mu\) is the corresponding hyperfine tensor between the NV center and that nucleus. For simplicity, we neglect the direct interaction between different nuclear spins at this stage, since it is typically much weaker than the NV-nuclear hyperfine coupling and does not affect the basic construction below.

\subsection{Interaction Engineering Framework}
To connect this system to our effective modulator-based picture, we restrict the NV center electronic spin to the two-level subspace spanned by \(\{|0\rangle,|-1\rangle\}\), and treat it as an effective qubit. We define \(\sigma_m^z = |{-1}\rangle\langle{-1}| - |0\rangle\langle 0|\). Accordingly,
\begin{equation}
    |{-1}\rangle\langle{-1}| = \frac{1+\sigma_m^z}{2}.
\end{equation}

Under a sufficiently strong static magnetic field and within the secular approximation, the coupling to the \(\mu\)-th $\Cs$ nuclear spin becomes
\begin{equation}
    \mH_\mu
    \approx
    \frac{\omega_\mu}{2} I_{\mu z}
    +
    |{-1}\rangle\langle{-1}|
    \left(
        A_{\mu\parallel} I_{\mu z}
        +
        A_{\mu x} I_{\mu x}
        +
        A_{\mu y} I_{\mu y}
    \right).
\end{equation}

This leads to a conditional interaction structure, which can be reshaped via control. Using microwave driving and dynamical decoupling, one can isolate a selected spin component and obtain
\begin{equation}
    \mH_{m\mu}^{(\alpha)}
    =
    \lambda_\mu^{(\alpha)} \sigma_m^z I_{\mu}^{\alpha},
    \qquad
    \alpha = x,y,z.
\end{equation}

We divide the $\Cs$ nuclei into battery spins \(b_i\) and charger spins \(c_k\), and define
\begin{equation}
    \tau_i^\alpha = 2 I_{b_i}^\alpha,
    \qquad
    \nu_k^\alpha = 2 I_{c_k}^\alpha.
\end{equation}

We assume two programmable interaction blocks,
\begin{equation}
    \mH_1
    =
    \sum_{i=1}^N g_i^{(B)} \sigma_m^x \tau_i^x,
    \qquad
    \mH_2
    =
    \sum_{k=1}^M g_k^{(C)} \sigma_m^y \nu_k^x,
\end{equation}
which serve as the building blocks for the engineered interactions.

\subsection{Modulator-Battery Coupling}
We first engineer the effective modulator-battery interaction. Starting from the interaction block \(\mH_1 = \sum_{i=1}^N g_i^{(B)} \sigma_m^x \tau_i^x\), we apply continuous microwave driving on the NV modulator and matching radio-frequency dressing on the selected battery nuclei. By moving to the rotating frame defined by the drives, the dominant interaction reduces to an exchange-type coupling. More explicitly, for the \(i\)-th battery spin, the effective Hamiltonian takes the form
\begin{equation}
    \mH_{mb}^{(i)}
    \approx
    \frac{\delta_m}{2}\sigma_m^z
    +
    \frac{\delta_{b_i}}{2}\tau_i^z
    +
    J_i
    \left(
        \sigma_m^+ \tau_i^-
        +
        \sigma_m^- \tau_i^+
    \right),
\end{equation}
where \(\delta_m\) and \(\delta_{b_i}\) are detunings with respect to the corresponding drive frequencies, and \(J_i\) is the effective flip-flop coupling strength determined by the transverse hyperfine interaction and the applied control fields.

Summing over all selected battery spins, we obtain
\begin{equation}
    \mH_{mB}^{\mathrm{eff}}
    \approx
    \frac{\delta_m}{2}\sigma_m^z
    +
    \sum_{i=1}^N
    \left[
        \frac{\delta_{b_i}}{2}\tau_i^z
        +
        J_i
        \left(
            \sigma_m^+ \tau_i^-
            +
            \sigma_m^- \tau_i^+
        \right)
    \right].
\end{equation}

In the approximately homogeneous regime, where the selected battery spins have similar parameters, we take
\begin{equation}
    J_i \approx J,
    \qquad
    \delta_{b_i} \approx \delta_b,
\end{equation}
and define the effective frequencies by \(\delta_m \approx -\omega_0\), \(\delta_b \approx -\omega_0\). The Hamiltonian then becomes
\begin{equation}
    \mH_{mB}^{\mathrm{eff}}
    \approx
    -\frac{\omega_0}{2}\sigma_m^z
    -
    \sum_{i=1}^N
    \frac{\omega_0}{2}\tau_i^z
    +
    \frac{J}{2}
    \sum_{i=1}^N
    \left(
        \sigma_m^x \tau_i^x
        +
        \sigma_m^y \tau_i^y
    \right).
\end{equation}

\subsection{Battery-Charger Coupling}
We next engineer the effective battery-charger interaction. This interaction is not directly available at the native level, and is instead generated perturbatively through Floquet commutator engineering. Starting from the two programmable blocks
\begin{equation}
    \mH_1
    =
    \sum_{i=1}^N g_i^{(B)} \sigma_m^x \tau_i^x,
    \qquad
    \mH_2
    =
    \sum_{k=1}^M g_k^{(C)} \sigma_m^y \nu_k^x,
\end{equation}
we consider the short Floquet cycle
\begin{equation}
    U_F
    =
    e^{-i\mH_1 \delta t}
    e^{-i\mH_2 \delta t}
    e^{+i\mH_1 \delta t}
    e^{+i\mH_2 \delta t},
\end{equation}
where \(\delta t\) is a short control interval. Using the Baker-Campbell-Hausdorff (BCH) expansion, the leading-order effective Hamiltonian generated over one cycle is proportional to the commutator
\begin{equation}
    \mH_{BC}^{\mathrm{eff}}
    \propto
    i[\mH_1,\mH_2].
\end{equation}
Evaluating the commutator gives
\begin{equation}
    [\mH_1,\mH_2]
    =
    2i\,\sigma_m^z
    \sum_{i=1}^N
    \sum_{k=1}^M
    g_i^{(B)} g_k^{(C)}
    \tau_i^x \nu_k^x.
\end{equation}
Thus, the engineered interaction takes the form
\begin{equation}
    \mH_{BC}^{\mathrm{eff}}
    \propto
    \sigma_m^z
    \sum_{i=1}^N
    \sum_{k=1}^M
    \Lambda_{ik}\,
    \tau_i^x \nu_k^x,
\end{equation}
with \(\Lambda_{ik} \sim g_i^{(B)} g_k^{(C)} \delta t\).

If the NV modulator is prepared and stabilized in a definite \(\sigma_m^z\) sector during this stage, then \(\sigma_m^z\) can be replaced by its expectation value, yielding an induced nuclear-spin interaction
\begin{equation}
    \mH_{BC}^{\mathrm{ind}}
    =
    \sum_{i=1}^N
    \sum_{k=1}^M
    \Lambda_{ik}\,
    \tau_i^x \nu_k^x.
\end{equation}

Under the rotating-wave approximation (RWA), the counter-rotating terms are neglected, leaving
\begin{equation}
    \mH_{BC}^{\mathrm{RWA}}
    \approx
    \sum_{i=1}^N
    \sum_{k=1}^M
    \tilde{\Lambda}_{ik}
    \left(
        \tau_i^+ \nu_k^-
        +
        \tau_i^- \nu_k^+
    \right),
\end{equation}
where \(\tilde{\Lambda}_{ik}\) denotes the resonant effective coupling strength. This describes the coherent exchange of excitations between the battery spins and the charger-spin subset.

To recover the common charger mode appearing in our target Hamiltonian, we now treat the charger subset collectively. Assuming that the charger ensemble is polarized and remains in the low-excitation regime, its collective bright mode can be approximated as an effective bosonic degree of freedom. 

The effective coupling matrix factorizes as $\tilde{\Lambda}_{ik} = \lambda_i \eta_k$, we define the collective bright-mode operators
\begin{equation}
    B^- =
    \frac{1}{\sqrt{\mathcal N}}
    \sum_{k=1}^M
    \eta_k \nu_k^-,
    \qquad
    B^+ =
    \frac{1}{\sqrt{\mathcal N}}
    \sum_{k=1}^M
    \eta_k^* \nu_k^+,
\end{equation}
with the renormalization factor
\begin{equation}
    \mathcal N = \sum_{k=1}^M |\eta_k|^2.
\end{equation}
Then the exchange term can be rewritten as
\begin{equation}
    \sum_{k=1}^M
    \tilde{\Lambda}_{ik}\,
    \tau_i^+ \nu_k^-
    =
    G_i \tau_i^+ B^-,
    \qquad
    G_i = \lambda_i \sqrt{\mathcal N}.
\end{equation}
In the approximately homogeneous regime, where the selected battery spins experience similar effective couplings to the collective charger mode, one may further take $G_i \approx g$, so that the battery-charger sector reduces to the uniform form assumed in the main text.

Accordingly, the battery-charger interaction becomes
\begin{equation}
    \mH_{BC}^{\mathrm{RWA}}
    \approx
    \sum_{i=1}^N
    g
    \left(
        \tau_i^+ B^-
        +
        \tau_i^- B^+
    \right).
\end{equation}

In the low-excitation limit, the collective charger mode may be treated using the Holstein-Primakoff approximation, under which
\begin{equation}
    B^- \approx b,
    \qquad
    B^+ \approx b^\dagger,
    \qquad
    [b,b^\dagger] = 1.
\end{equation}
The interaction is therefore reduced to
\begin{equation}
    \mH_{BC}^{\mathrm{bos}}
    \approx
    \sum_{i=1}^N
    g\left(
        \tau_i^+ b
        +
        \tau_i^- b^\dagger
    \right),
\end{equation}
which is precisely the desired battery-charger coupling in the target model.

The collective charger mode also carries a free-evolution term in the rotating frame,
\begin{equation}
    \mH_C
    =
    \omega_1 b^\dagger b,
\end{equation}
where \(\omega_1\) is the effective collective detuning of the bright mode. Accordingly, the charger sector is described by
\begin{equation}
    \mH_{BC}^{\mathrm{eff}}
    =
    \sum_{i=1}^N
    g\left(
        \tau_i^+ b+\tau_i^- b^\dagger
    \right)
    +
    \omega_1 b^\dagger b.
\end{equation}

This bosonic description is valid as long as the number of charger excitations remains much smaller than the total number \(M\) of charger spins. Beyond this regime, non-bosonic corrections become important. In addition, inhomogeneity in the charger subset can induce leakage from the bright mode into dark modes, which provides a natural limitation on the quality of the collective-mode approximation.

\subsection{Assembly of the Target Hamiltonian}

Collecting the above ingredients, we obtain two effective building blocks. The first is the modulator-battery sector,
\begin{equation}
    \mH_{mB}^{\mathrm{eff}}
    =
    -\frac{\omega_0}{2}\sigma_m^z
    -
    \sum_{i=1}^N
    \frac{\omega_0}{2}\tau_i^z
    +
    \frac{J}{2}
    \sum_{i=1}^N
    \left(
        \sigma_m^x\tau_i^x+\sigma_m^y\tau_i^y
    \right),
\end{equation}
while the second is the battery-charger sector,
\begin{equation}
    \mH_{BC}^{\mathrm{eff}}
    =
    \sum_{i=1}^N
    g
    \left(
        \tau_i^+ b+\tau_i^- b^\dagger
    \right)
    +
    \omega_1 b^\dagger b.
\end{equation}

These two sectors do not need to arise simultaneously from a single static control setting. Instead, they can be combined within a short-period Floquet protocol,
\begin{equation}
    U_F(T)
    =
    e^{-i \mH_{BC}^{\mathrm{eff}} \Delta t_{BC}}
    e^{-i \mH_{mB}^{\mathrm{eff}} \Delta t_{mB}},
    \quad
    T=\Delta t_{mB}+\Delta t_{BC}.
\end{equation}

In the high-frequency limit, the corresponding effective Hamiltonian is
\begin{equation}
    \mH_F
    \approx
    \frac{\Delta t_{mB}}{T}\mH_{mB}^{\mathrm{eff}}
    +
    \frac{\Delta t_{BC}}{T}\mH_{BC}^{\mathrm{eff}},
\end{equation}
up to higher-order Floquet corrections.

By tuning the effective couplings and the duty-cycle ratio, one reconstructs the desired many-battery Hamiltonian in the sense of effective simulation,
\begin{equation}
    \begin{split}
        \mH_F
        \approx
        &-\frac{\omega_0}{2}\sigma_m^z
        -
        \sum_{i=1}^N
        \frac{\omega_0}{2}\tau_i^z
        +\frac{J}{2}
        \sum_{i=1}^N
        \left(
            \sigma_m^x\tau_i^x+\sigma_m^y\tau_i^y
        \right)\\
        &+
        g\sum_{i=1}^N
        \left(
            \tau_i^+ b+\tau_i^- b^\dagger
        \right)
        +
        \omega_1 b^\dagger b.
    \end{split}
\end{equation}

This matches the target Hamiltonian of the main text, demonstrating that the proposed modulator-assisted many-battery model can be realized at the effective level in an NV-$\Cs$ platform.

\subsection{Remarks on coupling homogeneity}

Throughout the main text, we have adopted an approximately homogeneous effective description in which the modulator-battery couplings and battery-charger couplings are taken to be uniform, i.e. \(J_i \approx J\) and \(G_i \approx g\). This assumption is not expected to be exact in a realistic NV-$\Cs$ implementation, where the effective couplings generally depend on the microscopic spatial configuration and control selectivity of the individual nuclear spins.

The main role of this homogeneous approximation is to isolate the collective many-body charging physics in its clearest form and to preserve the permutation symmetry of the effective model. In particular, the symmetric structure allows the dynamics to be efficiently restricted to the corresponding collective subspace, which greatly simplifies both the analytical interpretation and the numerical simulation of the many-battery regime.

Importantly, coupling homogeneity is NOT essential to the physical mechanism proposed in this work. Moderate inhomogeneity in the effective couplings is expected primarily to broaden and quantitatively renormalize the charging dynamics, rather than to qualitatively alter the interaction structure or remove the collective charging channel altogether. In this sense, the assumption \(J_i \approx J\) and \(G_i \approx g\) should not be viewed as a strict experimental requirement. 

In particular, the charging-power enhancement discussed in the main text is expected to remain qualitatively robust against moderate departures from perfect homogeneity.

\subsection{Implementation of repeated modulator kicks}

In the charging protocol studied in the main text, the evolution is interspersed with repeated local kicks applied to the modulator. In the present NV-$\Cs$ platform, such kicks can be implemented directly on the NV electronic spin by short resonant microwave pulses.

At the effective level, the continuously engineered evolution is governed by the Floquet Hamiltonian \(\mH_F\), while each kick is represented by the unitary
\begin{equation}
    U_\text{kick} \propto e^{-i\frac{\pi}{2}\sigma_m^z},
\end{equation}
which is equivalent to \(\sigma_m^z\) up to an overall phase. Thus, one period of the kicked evolution takes the form
\begin{equation}
    U(\tau)
    =
    U_\text{kick} e^{-i \mH_F \tau}.
\end{equation}

Physically, this requires that the kick duration be much shorter than the effective Floquet timescale, so that the pulse may be treated as instantaneous relative to the engineered free evolution. Since the kick acts only on the NV modulator, it can be implemented without directly disturbing the battery and charger nuclear subsets. In this way, the repeated-\(\sigma_m^z\) kicking protocol of the main text can be naturally incorporated into the effective Hamiltonian simulation described above.

\subsection{Initial state preparation}

The relevant initial states required by the protocol can be prepared using standard NV-center control techniques. The NV electronic spin can be initialized optically into the \(|0\rangle\) state, which serves as a convenient reference state for the modulator qubit. The battery nuclei can then be prepared near their spin-down states by combining polarization transfer from the NV center with selective nuclear control, so that the battery sector starts close to its uncharged configuration.

For the charger subset, the natural reference state is a polarized low-excitation state. In the collective-mode picture, this corresponds to preparing the charger ensemble close to the vacuum of the bright mode \(b\). A finite initial charger excitation can then be created by transferring one or a few excitations from the NV center or from an auxiliary control step into the collective bright mode. In this way, one obtains the low-excitation initial states required for the Holstein-Primakoff description adopted above. For example, a single collective charger excitation corresponds to the bright-state preparation \(B^+|0\rangle_C\).

Therefore, the initial configuration assumed in the main text---namely, a modulator prepared in a definite computational state, battery spins near their ground states, and a charger prepared in a controlled low-excitation state---is compatible with standard initialization strategies in the NV-$\Cs$ platform.

\subsection{Timescale hierarchy and validity conditions}

The protocol described above involves two distinct time intervals, which play different physical roles. The first, denoted by \(\delta t_1\), is the short Floquet step used to assemble the effective Hamiltonian from the modulator-battery and battery-charger sectors. The second, denoted by \(\delta t_2\), is the physical interval between successive local kicks applied to the modulator qubit.

To implement the effective Hamiltonian stroboscopically, one may consider a short Floquet assembly step of the form
\begin{equation}
    U_F(\delta t_1)
    =
    e^{-i \mH_{BC}^{\mathrm{eff}} \delta t_1}
    e^{-i \mH_{mB}^{\mathrm{eff}} \delta t_1}.
\end{equation}
In the high-frequency limit, this realizes the desired combined evolution up to higher-order Trotter and Floquet corrections. Accordingly, \(\delta t_1\) must be chosen sufficiently short so that the noncommutativity of the two sectors remains perturbative. A natural condition is
\begin{equation}
    \|\mH_{mB}^{\mathrm{eff}}\| \delta t_1 \ll 1,
    \qquad
    \|\mH_{BC}^{\mathrm{eff}}\| \delta t_1 \ll 1.
\end{equation}
Equivalently, \(\delta t_1\) should remain much shorter than the inverse of the dominant effective energy scales of the problem, such as \(\omega_0^{-1}\), \(\omega_1^{-1}\), \(J^{-1}\), and \(g^{-1}\).

By contrast, the interval \(\delta t_2\) is not the Floquet assembly step, but the physical time between repeated modulator kicks in the charging protocol. At the effective level, the kicked evolution takes the form
\begin{equation}
    U(\delta t_2)
    =
    U_\text{kick} e^{-i \mH_F \delta t_2},
\end{equation}
where \(U_\text{kick} \propto \sigma_m^z\) is the kick operator and \(\mH_F\) is the effective Floquet Hamiltonian generated by the short assembly steps above. In this picture, \(\delta t_2\) is the protocol parameter that controls the kicking frequency and should therefore be chosen according to the dynamical regime of interest in the main text.

To justify the separation between the assembly dynamics and the kick protocol, one typically requires
\begin{equation}
    \delta t_1 \ll \delta t_2,
\end{equation}
so that many short Floquet assembly steps can be performed between two consecutive kicks. If the kick itself is implemented by a resonant microwave pulse of duration \(t_p\), then one further requires
\begin{equation}
    t_p \ll \delta t_2,
\end{equation}
so that each kick may be treated as effectively instantaneous relative to the engineered free evolution.

Taken together, these conditions define a hierarchy of timescales for the present implementation. In particular, the protocol is most cleanly realized when the Floquet assembly step is the shortest dynamical timescale of the effective simulation, while the kick interval \(\delta t_2\) remains short compared with the relevant coherent dynamics whenever one wishes to approach the dense-kicking or Zeno-like regime discussed in the main text.

\subsection{Noise and decoherence considerations}

In a realistic implementation, the effective coherent dynamics derived above will be modified by relaxation, dephasing, and collective-mode losses. At the level of the effective model, these effects can be incorporated using a Lindblad master equation of the form
\begin{equation}
    \begin{split}
        \dot{\rho}
    =&
    -i[\mH_F,\rho]
    +
    \gamma_{1,m}\mathcal D[\sigma_m^-]\rho
    +
    \gamma_{\phi,m}\mathcal D[\sigma_m^z]\rho\\
    &+
    \sum_{i=1}^N
    \left(
        \gamma_{1,b}\mathcal D[\tau_i^-]\rho
        +
        \gamma_{\phi,b}\mathcal D[\tau_i^z]\rho
    \right)
    +
    \kappa \mathcal D[b]\rho,
    \end{split}
\end{equation}
where
\begin{equation}
    \mathcal D[L]\rho
    =
    L\rho L^\dagger
    -
    \frac{1}{2}
    \{L^\dagger L,\rho\}.
\end{equation}
Here \(\gamma_{1,m}\) and \(\gamma_{\phi,m}\) denote the relaxation and dephasing rates of the NV modulator, \(\gamma_{1,b}\) and \(\gamma_{\phi,b}\) are the corresponding rates of the battery spins, and \(\kappa\) is the effective loss rate of the collective charger mode.

These noise channels affect different parts of the protocol in physically distinct ways. Decoherence of the NV modulator is particularly relevant because it directly degrades both the fidelity of the engineered modulator-battery interaction block and the repeated-kick control sequence. In contrast, battery-spin relaxation and dephasing primarily limit the amount of energy that can be stored and retained in the battery sector after charging. Finally, loss of excitations from the collective charger mode reduces the effective energy supply available for transfer into the battery subsystem.

Importantly, these effects do not immediately destroy the structure of the effective charging mechanism itself. Rather, they renormalize its quantitative performance. In particular, one expects a reduction in the achievable maximum stored energy, a suppression of the peak charging power, and a shortening of the storage lifetime after charging. However, the underlying interaction structure---namely, modulator-battery exchange, battery-charger exchange, and kick-controlled switching---remains complete at the effective level.

This identifies a natural parameter window for experimental observability. The protocol is expected to remain meaningful provided the relevant coherent interaction scales dominate over the corresponding loss and dephasing rates, namely
\begin{equation}
    \gamma_{1,m},\gamma_{\phi,m},\gamma_{1,b},\gamma_{\phi,b},\kappa\ll J,\ g,\ \omega_1.
\end{equation}

Therefore, even without performing a noise-level numerical analysis, the effective description already makes clear how the dominant imperfections enter the protocol and how they affect its performance. At this level, noise is expected primarily to reduce the achievable charging power, stored energy, and storage lifetime, rather than to change the interaction structure underlying the proposed modulator-assisted charging paradigm.

\end{document}